\documentclass[default,iicol]{sn-jnl}

\jyear{2021}%

\theoremstyle{thmstyleone}%
\theoremstyle{thmstyletwo}%

\theoremstyle{thmstylethree}%

\begin{document}

\title[Article Title]{Static and dynamic ordering of magnetic repelling particles under confinement: disks vs bars}

\author[1]{\fnm{M.} \sur{Aguilar-González}}

\author[1]{\fnm{L. F.} \sur{Elizondo-Aguilera}} 

\author[2]{\fnm{Y.D.} \sur{Sobral}}

\author*[1]{\fnm{F.} \sur{Pacheco-V\'azquez}}\email{fpacheco@ifuap.buap.mx}

\affil[1]{Instituto de F\'isica, Benem\'erita Universidad Aut\'onoma de Puebla, Apartado Postal J-48, Puebla 72570, Mexico}

\affil[2]{\orgdiv{Departamento de Matem\'atica}, \orgname{Universidade de Bras\'ilia}, \orgaddress{\street{Campus Universit\'ario Darcy Ribeiro}, \postcode{70910-900}, \state{Bras\'ilia-DF}, \country{Brazil}}}


\abstract{We explored experimentally the self-organization at rest and the compression dynamics of a two-dimensional array of magnetic repelling particles, using two particle geometries, namely, disks and rectangular bars. Despite the non-contact interaction, typical static features of granular materials are observed for both particle shapes: pile formation with an angle of repose and pressure saturation (Janssen-like effect), which can be explained by considering the magnetically-induced torques that generate friction between particles and confining walls. Particle shape effects are mainly observed during compression: while disks rearrange increasing the hexagonal ordering, bars augment their orientational ordering forming larger non-contact force chains; however, in both cases, the resistance to compression rises continuously, in contrast with the fluctuating compression dynamics (stick-slip motion or periodic oscillations) that characterizes granular systems with inter-particle contacts. The continuous response to compression, and the reduction of particle wear due to non-contact interactions, are desirable features in designing magnetic granular dampers.}

\keywords{Repelling grains, confined granular matter,  Janssen effect,  granular dampers.}



\maketitle

\clearpage

\section{Introduction}

Many damping systems employed across technological and practical applications leverage the distinctive properties of granular matter, in which the pervasiveness of energy dissipation allows for an efficient mitigation of mechanical stimuli, such as compressions, vibrations, impacts and seismic effects \cite{varela, braj, terzioglu1, hamzeh, terzioglu2,lu,Pacheco2013}. In such granular dampers --or particle dampers, which in essence are grains enclosed in a receptacle that is attached or embedded in a moving structure \cite{Sanchez2012}-- both the packing fraction and the particles characteristics (e.g., size, shape, roughness) play crucial roles in determining the energy dissipation processes. Consequently, these parameters are also significant in improving the efficiency of granular dampers \cite{terzioglu1,hamzeh}.  On the other hand, the behavior of granular matter under confinement is strongly influenced by inhomogeneous stress distributions \cite{rozenblat}, which stem from the multiple frictional contacts between the constitutive particles and, also, from the forces exerted between the grains and the confining walls, leading to intricate physical scenarios either at rest \cite{Azadeh2001, Jaeger1989,degennes,Pacheco2021} or under flow \cite{Beverloo1961, Hilton2011, Alvaro2012, Rubio2015, Arean2020}. Since the predominant interactions are highly dissipative, the physical description of the collective behavior in these non-equilibrium many-body systems is out of the reach of the standard tools of statistical mechanics, thus complicating their physical description and understanding.

While most of the aforementioned granular dampers are built using ordinary particles (sand, glass or metal beads, gravel, etc), there are other dissipative systems which might offer alternative routes to investigate the connection between energy dissipation and the response to mechanical stimuli. Such is the case of granular systems comprised by magnetic repelling particles under confinement \cite{opsomer}, in which the frictional contacts among particles are obliterated due to the magnetic repulsion. These systems have been the subject of different studies regarding their static and dynamic behavior \cite{thorens,Pacheco2015,lumay,Modesto2022,Cox,tsuchikusa}. 
For instance, for the case of repelling disks discharged under gravity from a two-dimensional cell, a constant flow rate was measured, with discharge profiles resembling closely those of conventional granular materials and, hence, can be reasonably described by a Beverloo-like correlation \cite{Beverloo1961,Pacheco2015}. As for the compression dynamics, recent experiments and simulations revealed that the increase of the compressive strength is linked to the development of spatial (hexagonal) ordering of the repelling disks \cite{Modesto2022}. Upon the addition of a second species, however, a system of bidisperse magnetic disks displays a more robust landscape for the compression response, now influenced not only by the total concentration of particles, but also by their size and mixing ratios, leading to different homogeneous spatial distributions of particles and, hence, to different scenarios for the compression response \cite{tsuchikusa}.  
In this context, a less explored route for tuning the compressive strength of a magnetic system (i.e., to improve the efficiency of a magnetic granular damper) is to modify the shape of the particles, thus giving them additional degrees of freedom associated to their rotations. Analyzing this possibility could provide valuable insight on both the theoretical modeling of repelling many-body systems and the practical usefulness of magnetic dampers. To the best of our knowledge, a systematic study on the static properties and the dynamic response to compression of a granular system comprised by anisotropic magnetic particles is still missing.

Motivated by the above reasons, in this work we report the results of an experimental study regarding the collective behavior of magnetic particles confined in a Hele-Shaw cell. As we show in what follows, despite the obliteration of frictional contact forces among the particles, our experimental systems display prototypical features of static granular matter, such as the development of an angle of repose and, more crucially, the Janssen effect \cite{degennes}. To highlight the influence of particle geometry on these features, we have considered and compared two types of magnetic particles, namely, disks and rectangular bars (both, with a magnetic moment oriented in the direction transverse to the cell walls to generate a repelling interaction).  For such magnetic granular systems, we also analyze the response under compression. Different from the axially symmetric case, in which the growth of the compressive strength is mainly due to the gradual spatial ordering of the disks, in a system of magnetic bars we found that essentially the same compression response is obtained in the absence of spatial ordering, as revealed by the analysis of the $\psi'_4(t)$ and $\psi'_6$ bond-orientational parameters \cite{Lopez2023}. Instead, our findings show that the compression strength is mediated by both the compression rate and the gradual orientational ordering of the bars, which tend to develop larger chain forces in comparison to the case of disks, and also, locally ordered domains that become increasingly percolated. Our results also indicate that a magnetic granular system can be approached simply as a continuum medium, in spite of the specific shape of the constitutive particles. 

Our work is organized as follows. In Section \ref{section2} we briefly describe the materials and methods employed to carry out the experiments. Section \ref{section3} describes our main results regarding the static behavior of a granular system of magnets. For clarity, however, this Section is divided in two Subsections, dealing with the development of an angle of repose (Subsection \ref{subsection3.1}) and the Janssen effect (Subsection \ref{subsection3.2}). In Section \ref{section4} we describe and discuss the response of the magnetic systems to compression, and highlight both the similitude and main differences between a system of disks and bars. Finally, in Section \ref{section5} we provide our main conclusions and perspectives. 

\section{Materials and methods}\label{section2} 

Our experimental setup builds upon prior work \cite{Pacheco2015, Modesto2022,tsuchikusa} and comprises a Hele-Shaw cell, filled with a  number $N$ of magnetic particles. The cell is composed of two parallel glass plates with length $L=150$ cm, width $w=39$ cm and spaced apart by a distance of $3.1$ mm. The dipolar moment of each of the particles is oriented nearly perpendicular to the glass plates. This orientation is crucial, as it allows to obtain non-contact repulsive interactions among the particles, which are obviously impeded from flipping due to the confinement of the glass plates.  To rule out frictional contributions from the lateral walls, additional magnetic bars were fixed to the edges of the cell. Therefore, the only frictional forces in our magnetic granular system are those involving the contacts between the particles and the confining walls, which are induced by the net torque ($\vec{\tau} = - \vec{\mu}_m \times \vec{B}$) acting on each particle due to the interaction of its magnetic dipole moment $\vec{\mu}_m$ with the magnetic field $\mathbf{B}$ of its neighbours (see details in Ref. \cite{Modesto2022}).

To investigate the influence of anisotropy in the magnetic repulsion, we have employed neodymium particles with two different shapes, namely, disks and rectangular bars. In the former case, we employed magnetic disks with mass $m=0.4$ g, diameter $d=5$ mm, thickness $t=3$ mm and magnetic field strength $H=10.08-10.4$ kG \cite{disks}; whereas in the later case, the particles considered had a mass $m=2.2$ g, length $h=15.24$ mm, width $w=6.604$ mm, thickness $t=2.794$ mm and magnetic field strength $H=8.4-9$ kG (according to the supplier \cite{bars}). 

\subsection{Data acquisition}

To analyze both static configurations and the response under compressions, the system was recorded using a high-speed camera Photron SA3, and the videos processed afterwards using standard particle tracking protocols to acquire positions and trajectories. To determine both the apparent weight of granular columns in the static case (see Subsection \ref{subsection3.2}) and also the response of the system under compressions (see Section \ref{section4}), we employed a digital force sensor Mark-10 DFG-355 and a C-Beam Linear Actuator with maximum travel distance of $40$ cm and a maximum speed of $5$ cm/s. Additional details depending on the specific (static or dynamic) study are provided below in the corresponding section.

\section{Results: Static behavior}\label{section3}

\subsection{Angle of repose}\label{subsection3.1}

One distinctive feature of conventional granular matter in repose is its ability to stack, which results from a complex interplay between multiple factors such as gravity, particle specifics (shape, friction coefficients, elasticity, etc), particle-particle direct interactions, internal structure, and also external conditions, with all of them contributing to the energy dissipation that leads to a stacked granular material. A simple bulk parameter that characterizes globally the stacking of granular assemblies is the so-called angle of repose $\theta_R$, which, in some specific contexts is defined as the maximum angle at which grains can be piled without slumping or flowing \cite{hamza2018}. This quantity, thus, parameterizes the static configurations that derive from the balance between weight and static friction \cite{elekes,khanal}. 

In conventional granular materials, the non-uniform stress distributions are dispersed along intricate force chains that depend explicitly on the particle contacts and packing arrangements, and these distributions determine, in turn, the energy losses due to friction. As just discussed, however, the only frictional contributions in our present system are due to the contacts between the magnetic particles (which are prone to flip) and the confining walls (which constrain the dipole inversion). Hence, it is interesting to analyze the process of stacking for these particular conditions for energy dissipation.

\begin{figure}
\begin{center}
 \includegraphics[width=1.04\columnwidth]{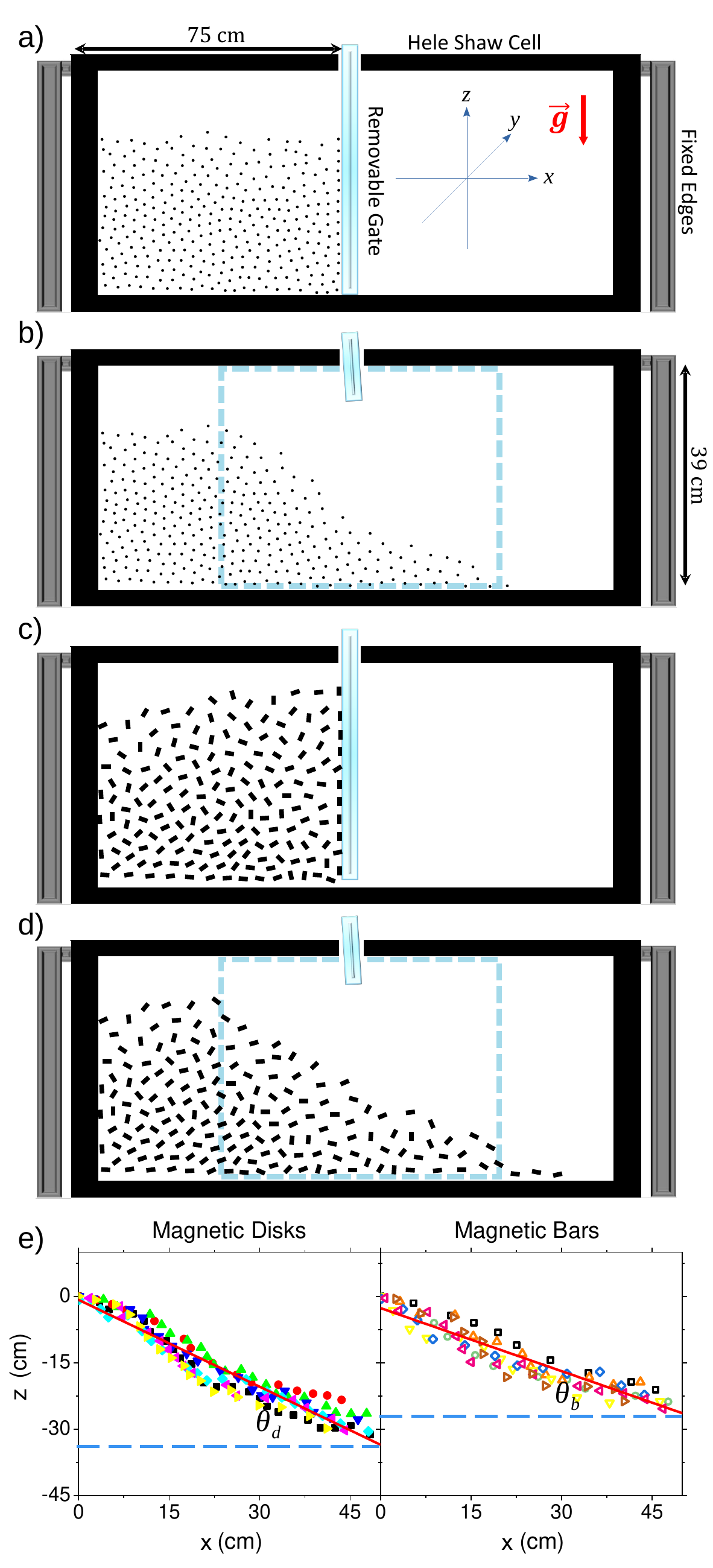}
 \caption{a-d) Setup used to obtain the angle of repose for magnetic disk and bars. The cell was divided in two identical sections using a removable gate. Initial/final configurations for a/b) disks and c/d) bars are shown, respectively. e) Surface layer profiles obtained from seven avalanche processes of the magnetic particles of each geometry, illustrating the development of an angle of repose for both disks (left panel) and rectangular bars (right panel).}
 \label{Figure1}
\end{center}
\end{figure}

To determine the existence of an angle of repose in our magnetic system we proceeded as follows: the (rectangular) cell was disposed with its width $w$ lying perpendicular to the horizontal plane, and its inner space was divided into two identical sections by a removable vertical gate. Then, one of the sub-cells was filled with either $N_d=532$ disks (see Fig. \ref{Figure1}a) or $N_b=250$ bars (see Fig. \ref{Figure1}c). After placing the magnets, the gate was quickly withdrawn vertically from the top of the cell to allow the magnetic particles to move in avalanche under the action of gravity, until they reached a new static configuration. The resulting particle distributions were registered with the camera and, from the profile of the surface layer in the window highlighted by the dashed squares in Figs. \ref{Figure1}b-d, the angle of repose was determined. The square window was chosen considering the region where the particles change their positions during the avalanche forming an inclined surface.

Fig. \ref{Figure1}e considers different profiles (symbols) obtained from seven realizations, for both disks (left panel) and bars (right panel). The first thing to notice in these results is the good degree of reproducibility for the two types of particles. In all cases, the profile of the surface layer shows a monotonically decreasing height (from left to right) thus allowing to estimate an average angle with respect to the horizontal $x$-axis. For reference, the average inclinations are highlighted in the two panels of Fig. \ref{Figure1}(e) by the solid lines. Overall, one observes that the magnetic disks display a larger angle of repose, $\theta^d_R=34^\circ$, with respect to that obtained for the bars, $\theta^b_R=26^\circ$, indicating that sharper granular heaps are obtained in the former case. This might be unexpected considering that, for the case of non-magnetic particles, elongated geometries are expected to form more inclined slopes \cite{khanal}. In our case, as we will discuss later, the repelling magnetic interaction avoids particle interlocking for the case of bars and promotes hexagonal ordering for the case of disks; which plays in favor of more stable configurations (larger angle of repose) for the latter case. For the case of disks, the average values obtained for the angle of repose are quantitatively comparable to those typically observed in conventional granular materials \cite{hamza2018}.

\begin{figure}[ht!]
\begin{center}
\includegraphics[width=\columnwidth]{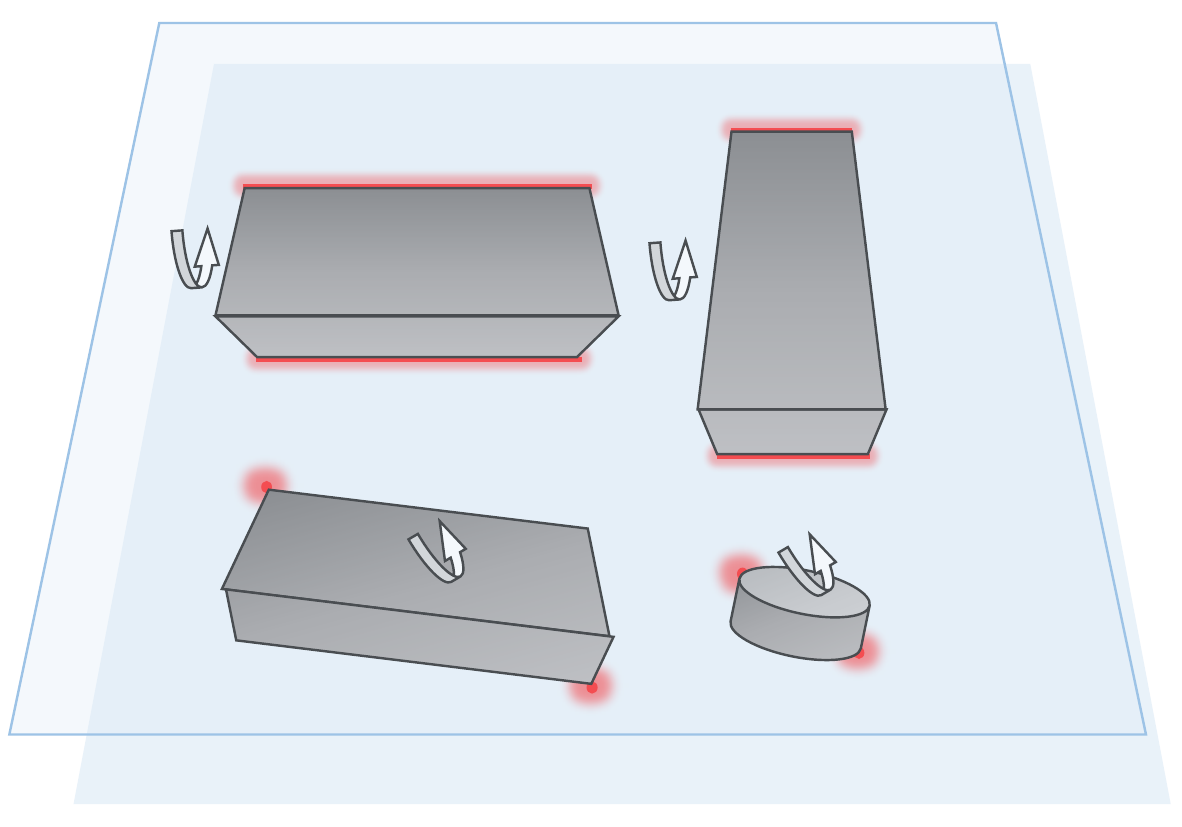}
 \caption{Possible contact zones (marked in red) between the bars or disks with the confining glass walls due to torques induced by the net magnetic field acting on each magnetic dipole. These contact points are the main contribution to friction in the system.}
 \label{Figure2}
\end{center}
\end{figure}

A few remarks might be in order here. Small particle numbers are limited by the size of the cell because one must keep in mind that, in reality, each magnetic particle corresponds to a larger repulsive core as compared to its physical dimensions. If we define the two-dimensional packing fraction $\phi_i$ as the ratio of the lateral area of $N_i$ particles ($i=b,d$) to the total area occupied by the stack inside the sub-cell, from image analysis one estimates $\phi_d \approx 0.05$ and $\phi_b \approx 0.1$, for disks and bars, respectively. Note that these values are one order of magnitude lower than typical loose packing fraction for non-repelling disks ($\phi \sim 0.60$) due to the repulsive nature of the particles. Of course, such larger repulsive shells suggest the formation of a network of non-contact repulsive forces among all the particles. These repulsive forces, in turn, determine both the spatial dispersion of the particles inside the cell, and also the stress distribution in the system, with the later being directed towards the confining walls. For the specific case of the bars, in addition, such repulsive cores of magnetic origin must be clearly anisotropic, leading thus to less stable static configurations that are sensitive to the relative orientations among the particles. These orientations also determine the axis around which each particle aims to flip and, hence, the contact surface between them and the glass plates, which at the end is responsible for the energy dissipation and maintains the stability of the stack. Note that depending on the particle geometry, friction can act in opposite points of the disk faces, or in opposite corners or sides of the bars, as it is sketched in Fig. \ref{Figure2}. In general, during any avalanche process, the bars are prone to reorient in the plane until a new static configuration is achieved. This ability of the conforming particles to reconfigure their frictional contributions through rotations enhances the tendency to flow, as compared to the case of disks, thus leading to smaller angles of repose.

\subsection{Janssen Effect}\label{subsection3.2}

Another distinctive feature of granular matter refers to the so-called Janssen effect \cite{sperl2,Janssen}). This is a long-known experimental fact: when grains are poured into a container, the weight measured at the bottom of the column eventually becomes smaller than the total weight of the grains \cite{hagen,sperl1}. Equivalently, the pressure within the material due to gravitational compaction does not increase linearly as the granular column becomes higher. Instead, it reaches a saturation value. This occurs because an increasingly larger fraction of the weight of the added mass is redirected towards the container walls through internal force chains formed by frictional contacts between particles \cite{degennes,Vanel1999,Vanel2000,Bratberg2005,Huang2016}.

As discussed in the previous section, our present system shows the formation of an angle of repose due to both the development of an inhomogeneous network of non-contact repulsive forces among the particles and, more crucially, the frictional forces that are exerted towards the walls via the magnetic dipolar torques. The question that naturally arises, therefore, is whether Janssen saturation also occurs in our system of repelling particles under two-dimensional confinement. Let us mention here that other manifestations of the Janssen effect in magnetic granular materials have been reported recently, although considering the application of external fields to magnetic particles under three-dimensional confinement \cite{thorens}. 

\begin{figure}[ht!]
\begin{center}
 \includegraphics[width=\columnwidth]{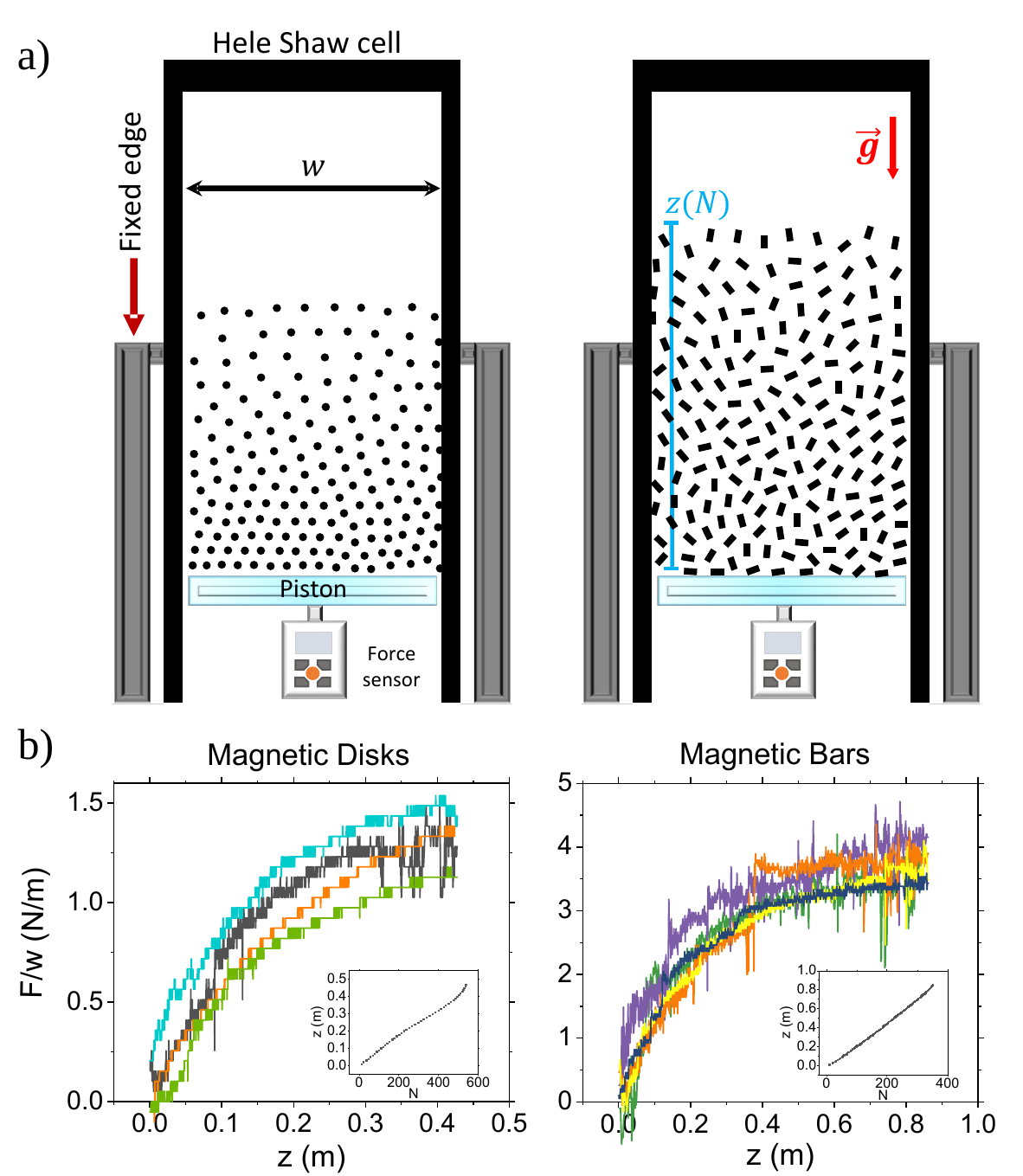}
 \caption{a) Schematic representation of granular columns obtained with magnetic disks (left panel) and magnetic bars (right panel) poured gently into the Hele-Shaw cell. b) Plots of the apparent weight of the granular columns $F$ (per width length $w$) formed with disks (left panel) and bars (right panel) versus the column's height $z(N_i)$, measured during four pouring processes. Although $z(N_i)$ is linearly proportional to the number of particles $N_i$ (see insets), the Janssen-like saturation is found for $F/w$ vs $z$ for both geometries.} 
 \label{Figure3}
 \end{center}
\end{figure}

To address the above question, we have disposed the Hele-Shaw cell now with its length oriented in the direction of gravity, and filled it gently with either magnetic disks or magnetic bars to obtain static columns with a nearly uniform top layer. For reference,  we show a schematic representation of our experimental setup in Fig. \ref{Figure3}a. In the case of disks we used $N_d=532$ magnets (total weight $\sim$ 2.1 N), whereas for rectangular bars we employed a number of particles $N_b=350$ (total weight $\sim$ 7.5 N), with both corresponding to the maximum number of magnetic particles within our experimental availability. At the bottom of the cell, a piston of length $w$ was coupled to a force sensor (Mark-10 DFG-355) to determine the force acting on the piston $F$ due to the column as a function of the number of particles $N_i$ ($i=d,b$) or, equivalently, the two-dimensional pressure $F/w$. 

Fig. \ref{Figure3}b displays the results obtained for different columns of either disks (left panel) or bars (right panel), summarized in plots of $F/w$ versus the height of the column $z(N_i)$. Let us mention here that, in all cases, the particles were poured gently into the cell in order to ensure that the spatial (and orientational) configurations of the columns were not altered significantly during the pouring process, and also, to obtain surface layers with particles located nearly at the same height. Following this procedure, the columns height $z$ increases almost linearly with the number of particles $N_i$ ($i=d, b$) (see the insets of both panels).   Unlike the results obtained for the angle of repose, here we found a poorer reproducibility across repeated trials, but this is also common for the static Janssen effect with typical granular material \cite{Katsuragi2016,Yann2003} (the reproducibility can be enhanced considering a system with moving walls \cite{Yann2003,Windows2019}). However, for all cases, we found a linear growth of the two-dimensional pressure with $z$ (small $N$), but the pressure profile $F/w$ bends over and gradually develops a plateau, thus indicating saturation. Notice that this behavior is more pronounced for the case of the bars, since we employed a number of particles carrying a greater amount of neodymium mass, but in the two types of particles the onset of saturation is clear. Therefore, as $z$ increases, the magnetic torques directed towards the walls generate enough friction to eventually support the weight of the column. The obtained curves of $F(z)$ resemble the pressure profile predicted by Janssen's continuum model for a conventional granular system \cite{Janssen,sperl2,degennes}. Hence, we may attempt to fit our data using a Janssen-like expression for a 2D friction-driven system. Following Ref. \cite{Karim2014}, for instance, we might consider the fitting function:

\begin{equation}
\frac{F}{w}=\mu_{g}\rho_i g\lambda_i (1-e^{-z(N_i)/\lambda_i}),
\label{equation1}
\end{equation}

\noindent in which $\mu_g\approx0.27$ is the friction coefficient between a neodymium particle and a glass plate (here determined from sliding experiments on an inclined plane), $\rho_i$ the mass density in a column with height $z(N_i)$, and $\lambda_i$ a characteristic length, such that $F/w= \mu_g \rho g \lambda_i$ for $z(N_i)\to\infty$. Let us highlight that, in conventional granular systems, this length scale is typically comparable to $w$.

\begin{figure*}[ht!]
\begin{center}
\includegraphics[width=10cm]{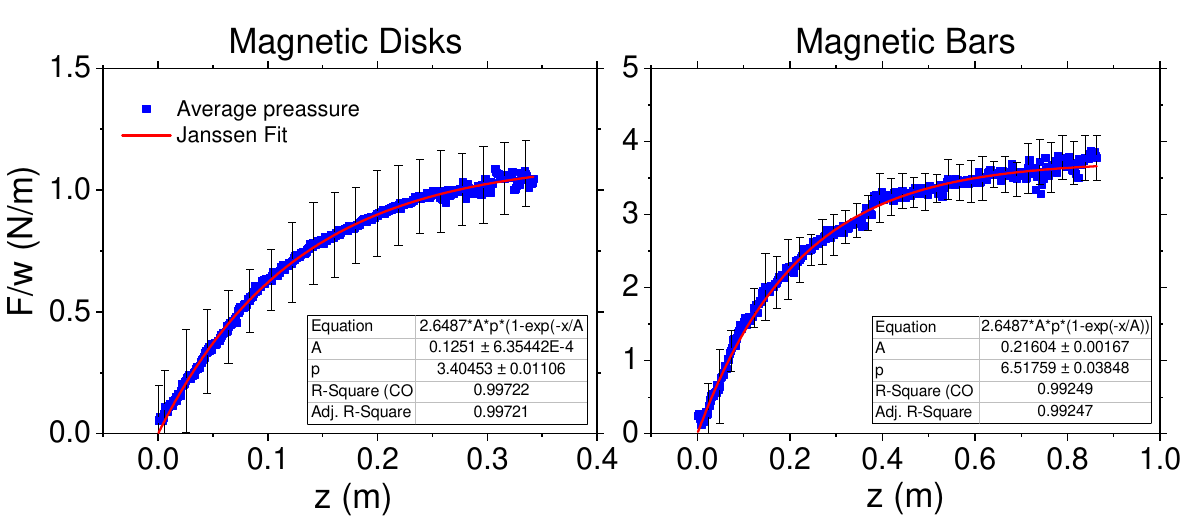}
 \caption{Plots of $F/w$ vs $z$ for disks (left panel) and bars (right panel). Blue symbols correspond to the average experimental values from the four experiments shown in Fig. \ref{Figure3}b for each geometry with the corresponding standard deviation, and red lines to the data fit using Eq. \eqref{equation1}.} \label{Figure4}
\end{center}
\end{figure*}

Fig. \ref{Figure4} considers the average pressure profiles (solid symbols) obtained from the results of Fig. \ref{Figure3}, along with their corresponding standard deviation (vertical bars). In the same figure, we also plot Eq. \eqref{equation1} using $\rho_i$ and $\lambda_i$ as fitting parameters. Such exercise, which shows a remarkable correspondence with the experimental data, yields the values $\rho_d=3.4$ kg/m$^2$, $\rho_b=6.5~\rm{kg/m^2}$,  $\lambda_d=0.125$ m, and $\lambda_b=0.265$ m. Remarkably, we could also determine the real mass densities separately, from $\rho_i=N_im_i/z(N_i) w$, which yields the values  $\rho_d=1.13$ kg/m$^2$ and $\rho_b=2.32$ kg/m$^2$, approximately one third of the value resulting from the fit. We attribute these quantitative differences to the distinct nature of the repulsive magnetic particles, whose effective excluded area is larger than the actual area of the particles. Let us recall that Eq. \eqref{equation1} is based on a continuum model analysis \cite{Janssen,deGennes1999}, and does not distinguish the nature of the constitutive particles. Moreover, one important assumption in deriving Eq. \eqref{equation1} for conventional granular matter is that the horizontal components of the stress tensor in the granular medium ($\sigma_{xx}$, $\sigma_{yy}$) are proportional to the vertical stress $\sigma_{zz}$. Based on the previous results, we conjecture that this assumption also holds in the present case, since the torque experienced by each particle redirects the magnetic pressure as a friction towards the frontal and rear walls. This would explain why Eq. \eqref{equation1} describes quite well the average saturation pressure found in our measurements.

In summary, all the above findings support the interpretation of a system of magnetic repelling particles as an effective granular material which, in many regards, can also be approached as a continuum medium regardless the geometrical characteristics of the constitutive particles. It is important to remark again that all the static features outlined above, which resemble qualitatively those found in conventional granular materials, emerge from the friction against the confining walls induced by the magnetic interaction among the particles. Since the macroscopic behavior of most granular systems is governed by the formation of contact force chains, it is natural to question the role of these kind of forces in determining the behavior of the system under consideration. To explore this further, we now turn our attention to the configurational and dynamical aspects of the system when it is subjected to compression. 

\section{Dynamic response to compression}\label{section4}

In recent years, the response of granular matter to mechanical stress has been the subject of considerable research, due to its significance in many practical applications, such as the development of increasingly sophisticated dampers and, more generally, in the physical characterization of reference model systems \cite{varela,braj,terzioglu1,hamzeh,terzioglu2,lu,Pacheco2013,Sanchez2012,rozenblat}. For ordinary granular materials, two qualitatively different compression responses can be observed. For systems comprised of rigid particles, the force required to compress the granular medium shows an exponential growth by increasing the compression length, with the occurrence of sudden drops that result from the spontaneous re-configuration of particles (stick-slip motion) \cite{Pacheco2021}. For brittle materials, instead, the resistance to compression grows as a power law by increasing the applied stress \cite{Pacheco2021-2}. In this case, one also observes a periodical formation of "propagation bands" in the medium, arising from multiple rupture events of particles at the bottom, and leading to periodical fluctuations on the increasing compressive force \cite{Valdes2012, Guillard2015}. 

The compression of systems conformed by repelling magnets, along with their relaxation during the subsequent expansion, is remarkably different to the above cases and has been investigated both experimentally and numerically \cite{Modesto2022,tsuchikusa,Cox}. However, this has been analyzed only for the case of cylindrical particles under quasi two-dimensional confinement, exhibiting a continuous exponential growth of the required compressive force with increasing compaction, but without displaying sudden drops or periodic oscillations. Such magnetic granular systems, in addition, show a history-independent relaxation during the expansion process that follows the compression. Importantly, in monodisperse systems, a progressive hexagonal ordering of the particles is found in the vicinity of the compressing piston, which induces an inversion of the density profile with respect to initial state \cite{Modesto2022}. This continuous particle organization underlies the smooth increase of the resistance to compression.

\begin{figure}[ht!]
\begin{center}
 \includegraphics[width=\columnwidth]{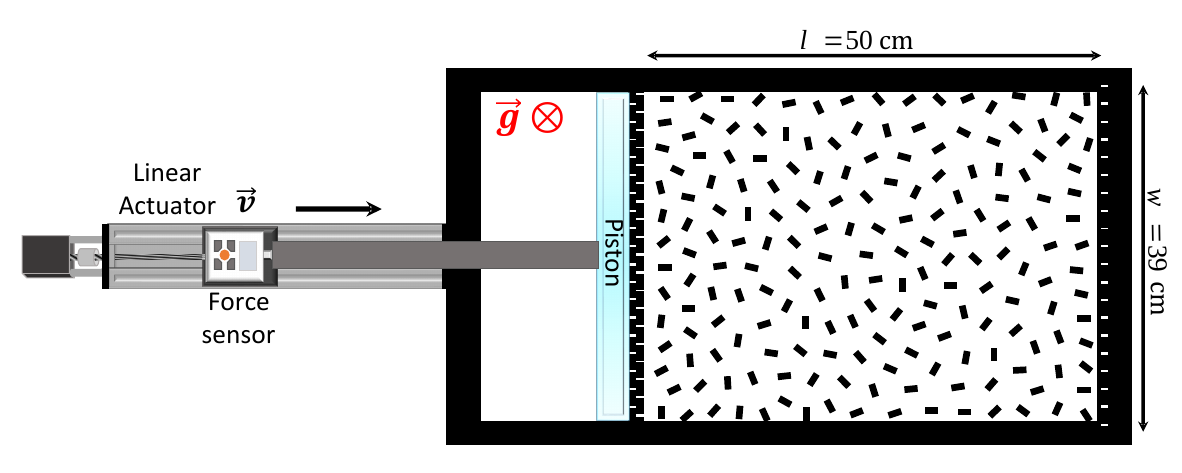}
 \caption{Schematic representation of the experimental setup used to measure the resistance to compression $F(x)$ opposed by the horizontal monolayer of magnetic particles (the inner space of the cell filled with bars is actually a photograph of the setup, used here to illustrate the initial state of the system).} \label{Figure5}
\end{center}
\end{figure}

To complement the above studies, we explored how the compressive response of a magnetic granular system depends on the geometry of its constituent particles. For the specific case of the magnetic bars, for example, the intrinsic anisotropy of the repulsive interactions implies two possible routes for particle ordering, namely, positional and orientational, which in turn must act in concert to mediate the response of the magnetic medium to an applied stress. 
To elucidate these aspects, we have conducted various compression experiments in samples consisting of either magnetic disks or bars. To disregard gravity-induced compaction, the Hele-Shaw cell was disposed horizontally (i.e. transversal to $\vec{g}$), allowing to focus solely on the influence of the anisotropy of the magnetic repulsion on the system's response to the compressions. As depicted in Fig. \ref{Figure5}, a piston with attached magnetic particles was employed to shrink gradually the samples towards different compression lengths $x_\text{{max}}$ in the horizontal plane (the sketch represents a top view of the cell). The piston was coupled to a C-Beam Linear Actuator, used to compact the samples occupying initially an area $A= l \times w$, at different compression rates $v$ and for different compaction ratios $\epsilon\equiv x_{\text{max}}/l$. In our experiments, we used the inner cell dimensions $l\approx 50~\rm{cm}$ and $w= 39$ cm for both geometries, thus yielding a total covered area of $A =1950$ cm$^2$. In all cases, a force sensor attached to the piston was employed to determine the force $F$ exerted by the magnetic system and opposing the compression. 

\begin{figure*}[ht!]
\begin{center}
\includegraphics[width=12 cm]{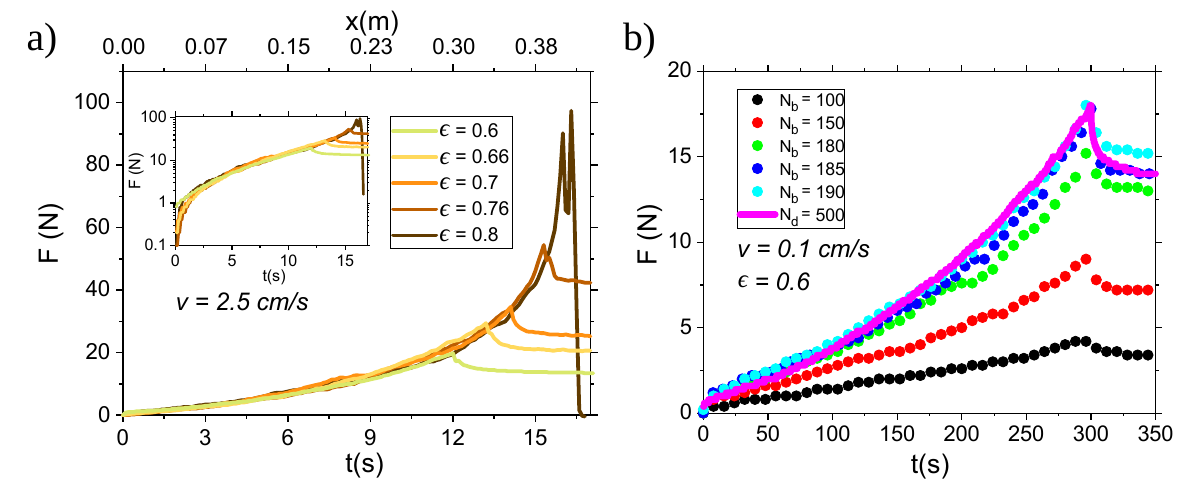}
 \caption{a) Force profile $F(t)$ (or, equivalently, $F(x)$) obtained for different compaction ratios $\epsilon$ and constant $v$ for the case of disks. The inset depicts the same graph in semilog, which suggests a nearly exponential trend in all the cases. b) Force profiles obtained by compressing samples constituted by $N_b$ magnetic bars (solid symbols), for a compression process with parameters $v=0.1$ cm/s and $\epsilon=0.6$. The solid line corresponds to the force profile obtained for $N_d=500$ disks for the same compression protocol. The relaxation in each case starts after 30 cm of compression ($t>300$ s).} \label{Figure6}
\end{center}
\end{figure*}

To serve as a reference for further discussion, in Fig. \ref{Figure6}a we display the results obtained for the response force $F$ in samples comprised by $N_d=500$ disks, compressed at fixed rate $v=2.5$ cm/s towards different compression ratios $\epsilon$, as indicated. Notice that the end of each process is marked by the onset of a relaxation in $F$. The first thing to highlight in these results is the absence of sudden drops in $F(t)$ during compression (or, equivalently in $F(x)$), despite the fact that multiple particle re-arrangements are observed over the process. This is in agreement with the results found for the same particle geometry under vertical compression \cite{Modesto2022}, indicating that gravity does not play a fundamental role in the particle rearrangement.  Hence, unlike a conventional granular system composed by rigid particles in contact, here we observe a continuous growth of the response force, obviously driven by the progressively stronger repulsion between magnetic particles that are brought closer and closer to each other by the piston.   Of course, this also leads to an increase in the pressure exerted towards the confining walls, given that the strength of the magnetic field acting on the dipolar moment of each magnet induces a torque-pair, but the dipole inversion is frustrated by the confining (rear and frontal) walls, thus increasing the friction experienced by each particle. 

One notices that, for all the compression ratios considered, the plot of $F$ versus time $t$ (or, equivalently, versus the compression length $x$) in Fig. \ref{Figure6}a follows a single master curve with a seemingly exponential growth. This exponential behavior is more clearly highlighted in the inset, which displays $F$ in a semilog scale. As already mentioned, for each selected value of $\epsilon$ a subsequent relaxation of the system occurs once compression ceases, marked by a sudden drop of $F$ that leads to a plateau value. Notice that such value becomes increasingly larger with greater compression ratio. Furthermore, for $\epsilon=0.8$ (corresponding to a compression length of $x_{\text{max}}\approx40$cm) we reach the maximum load supported by the walls of our Hele-Shaw cell, where the repelling disks are capable of overcoming the confining pressure and flip, collapsing the system abruptly and forming stacks of broken magnetic particles. For that reason, we limited our study to a maximum compression force $F < 30$ N.

With this context in mind, let us discuss now the most relevant features observed in the compressive response of the magnetic system upon replacing the repelling disks with bars. For this, one might be tempted at first glance to consider samples with the same amount of neodymium mass, but this time distributed in bar-shaped particles. This would require to consider only 90 of such particles, which yield approximately the same mass contained in 500 disks, namely, 200g. We observed, however, that this particle number leads to a remarkably weaker response force as compared to that obtained for the system of disks just described. Of course, this is expected from the fact that any given neodymium mass, partitioned into increasingly smaller portions, provides more contributions to the repulsive forces among the resulting pieces, so that the magnetic repulsion contributes more efficiently to the responsive force of the system (at fixed mass) by employing a larger number of smaller particles. 

Therefore, in order to obtain a system composed by repelling bars with a response comparable to that of a system conformed by 500 disks, we compressed samples with different number of bar-shaped magnets, and used the resulting force profiles to select an adequate value for $N_b$. This procedure is illustrated in Fig. \ref{Figure6}b, which displays the results obtained for the responsive force $F$ in samples with different $N_b$, as indicated, compressed at $v=0.1$ cm/s and for fixed compaction ratio $\epsilon=0.6$ (solid symbols). Using this information, thus, we choose empirically a number $N_b=185$ (equivalent to 407g of neodymium) as a proxy to compare with the system of 500 disks (solid line), since both yield essentially the same profile for the compressive force $F$, including its further relaxation, for given $v$ and $\epsilon$. 

\begin{figure}[ht!]
\begin{center}
 \includegraphics[width=\columnwidth]{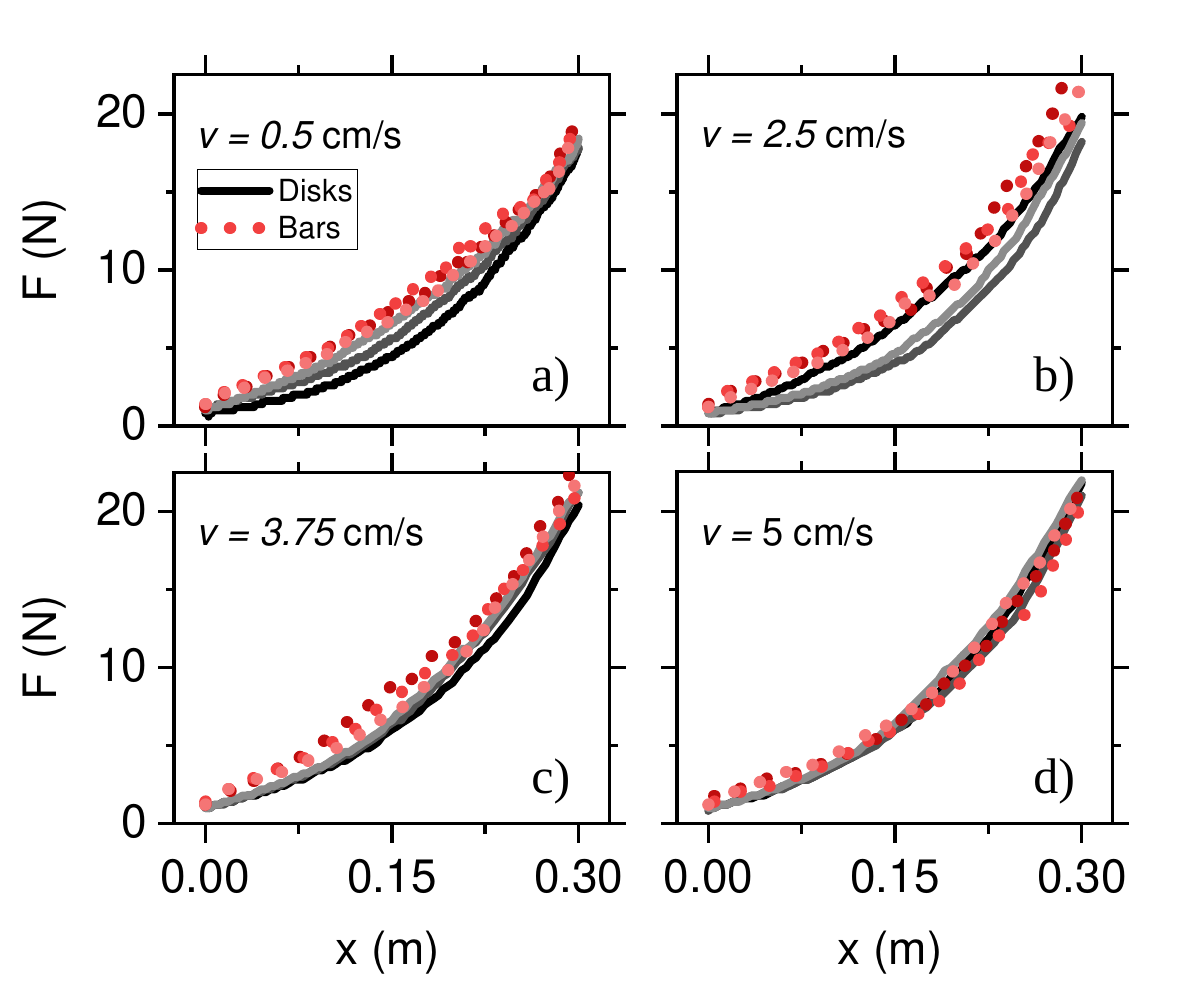}
 \caption{Force acting on the piston $F$  as a function of the compression length $x$ for different compression rates $v$ and for the two geometries. Same colored scale differentiates three independent realizations for the same geometry.} 
 \label{Figure7}
\end{center}
\end{figure}

We then analyzed the influence of the compression rate $v$ on the response $F$ of both systems. For this, we compacted samples with either $N_d=500$ disks or $N_b=185$ bars, at different rates $v$ and for fixed $\epsilon=0.6$ ($x_{max}=30$ cm). Fig. \ref{Figure7} summarizes the results obtained for disks (solid lines) and bars (symbols). At small compression rates, one observes similar exponential growths in $F$ upon increasing the compression length (Fig. \ref{Figure7}a), with the bars exhibiting a better reproducibility of the results, although both systems lead to the same cumulative force at the end of the compression. Of course, during each process the particles always undergo progressive changes in their spatial distribution, but particle re-arrangements are more prominently observed at small $v$. Larger displacements, however, were observed for the repelling disks, thus enabling them to explore a wider range of different configurations. The magnetic bars, in contrast, exhibited less mobility, along with a tendency to re-orient locally in specific directions, which we attribute to the anisotropy of their magnetic repulsion (these aspects will be further discussed in subsection \ref{subsection4.1}). We believe that these features underlie the lower reproducibility observed for the disks.
One notices that, for relatively faster compressions (e.g. $v=2.5$ cm/s and $v=3.75$ cm/s), the compressive strength becomes larger in both systems, with the samples comprised by bars displaying larger values in $F$ as compared to those conformed by disks (Figs. \ref{Figure7}b and \ref{Figure7}c). The geometrical characteristics of the repelling particles, however, are practically erased by increasing $v$ further, where the responsive-force profiles of both systems eventually collapse into a single curve (Fig. \ref{Figure7}d). 

The aforementioned results indicate that the compressive force of the present system is not significantly affected by the geometry of the constituent particles. Instead, the key control parameters are the total (neodymium) mass and the number of repelling particles employed, with the latter accounting for the amount of repulsive contributions to the inter-particle forces, which, in turn, determine the compressive strength of the system. Note that the same responsive force produced by a given mass of neodymium distributed in bars, can be obtained with approximately half that mass, but distributed in smaller disks.

\begin{figure}[ht!]
\begin{center}
 \includegraphics[width=\columnwidth]{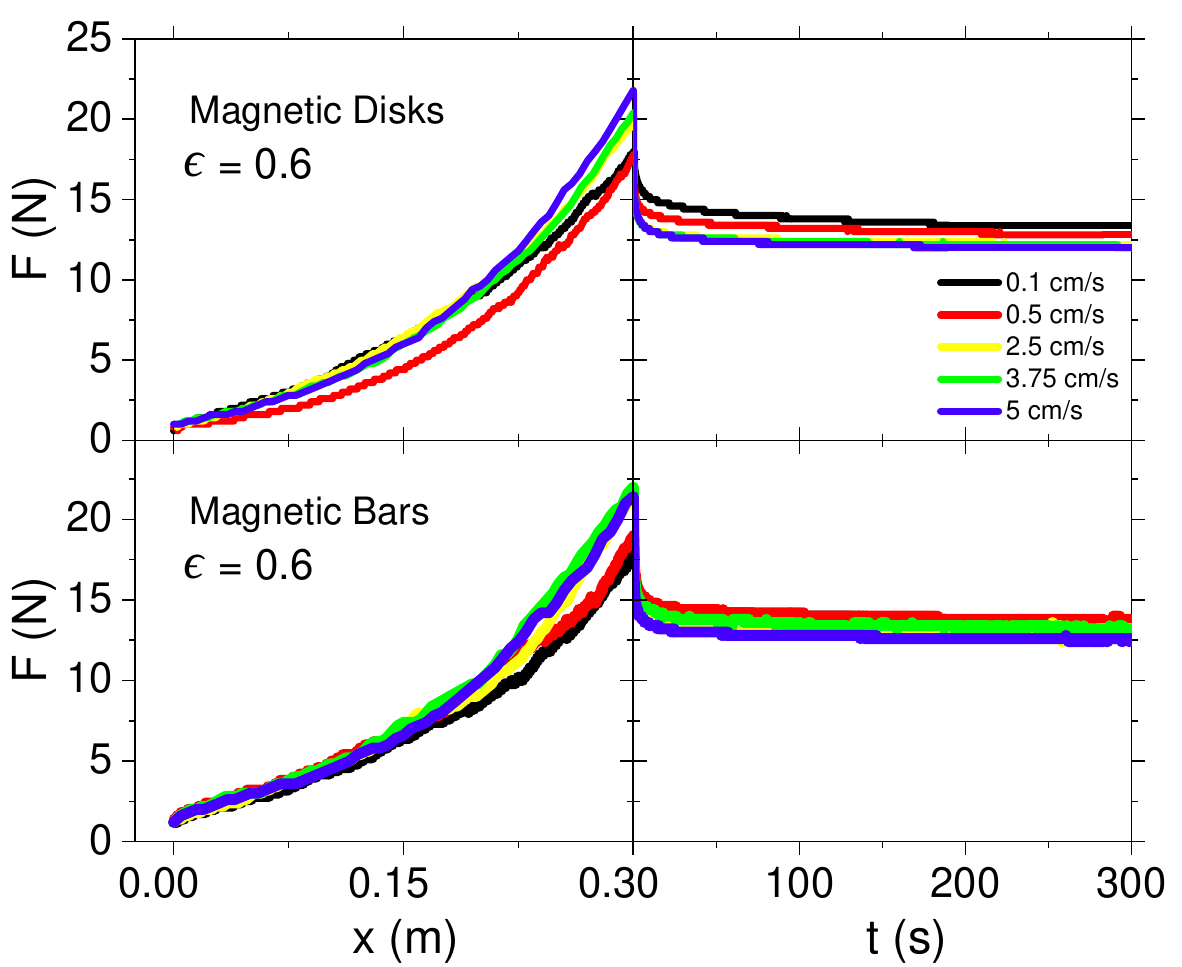}
 \caption{$F(x)$ during the compression phase measured up to reaching a maximum piston stroke $x_{max}=30 cm$. After reached this value, the relaxation process starts and $F$ is measured during 300 s. Experiments for different compression rates and for the two geometries are shown. Each line corresponds to one independent realization.} \label{Figure8}
\end{center}
\end{figure}

In the case of the bars, however, one intuitively expects that the anisotropy of the magnetic repulsion must lead to spatial distributions of particles that differ substantially from those obtained for the disks at the end of a compression process. This then raises the question of whether such configurational differences have any relevance on the relaxation behavior that follows a compression. Interestingly, we also found that such details do not play a role in determining the subsequent relaxation. This is emphasized in Fig. \ref{Figure8}, where we display a series of force profiles for both geometries, obtained for different compaction speeds (at fixed $\epsilon=0.6$), but this time also considering the evolution of $F$ after compression. Overall, this relaxation is shown to be almost independent of the particle geometry. Nonetheless, higher compression speeds lead to slightly higher resistance force, followed by a more pronounced force drop during the relaxation phase. Note, however, that the increase of $F$ is small even when the compression speed increases almost two orders of magnitude, but the relaxation to lower values is evident at faster compression rates, which has also been reported for compression-relaxation cycles of conventional rigid particles \cite{Pacheco2021}. At fast compression rates, the particles do not have time to reach the most stable configurations; therefore, upon the onset of relaxation, these particles experience small displacements towards more stable positions, which overall, results in a more pronounced relaxation in comparison to slower processes where the particles keep their relative positions after compression.
 
\subsection{Particle image velocimetry and orientational distributions during compression}\label{subsection4.1}

\begin{figure*}[ht!]
\begin{center}
 \includegraphics[width=\linewidth]{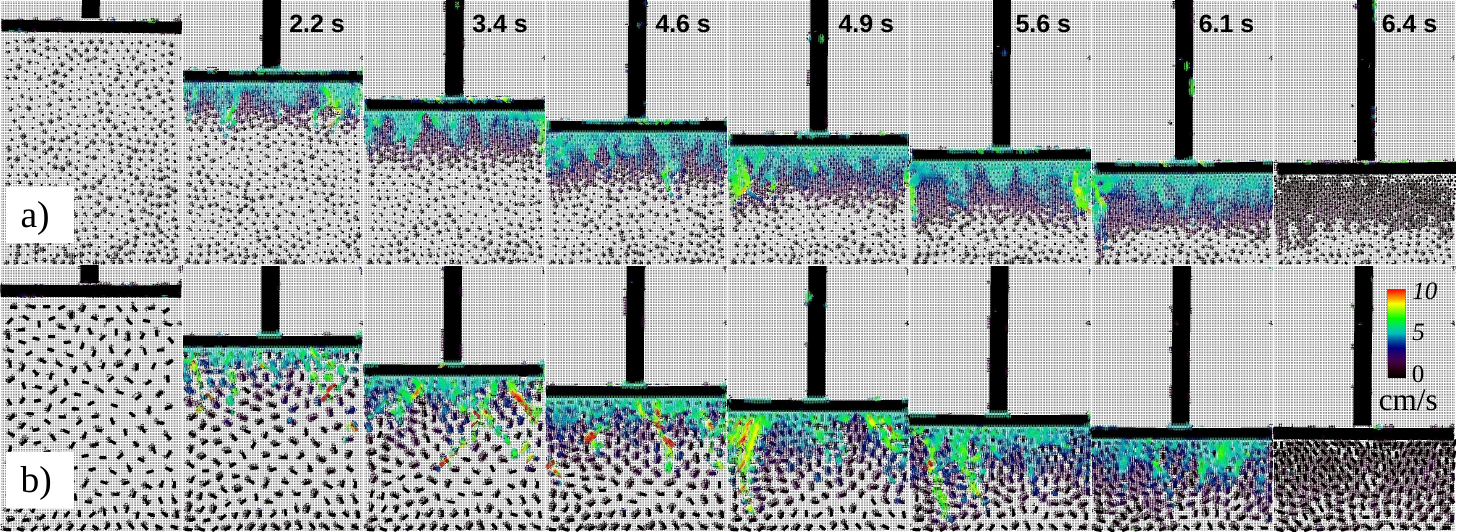}
 \caption{Top view snapshots showing the results of PIV analysis during the compression of the magnetic granular medium for the case of a) disks and b) bars, with the conditions $v = 5$ cm/s and $\epsilon=0.6$. The color scale indicates the corresponding particle velocity. The snapshots were rotated $90^{\circ}$ for better visualizing the relative piston displacement (actually, the motion is along the horizontal plane). }
 \label{Figure9}
\end{center}
\end{figure*}

To gain insight into the microscopic mechanisms underlying the above features, we also carried out particle image velocimetry (PIV), which allows to track the individual motion of the particles during compression. For this, we recorded the evolution of the system over the compaction process, and applied standard particle-tracking protocols to determine the individual trajectories of the particles. From a comparative analysis between consecutive images, we were able to determine the particles' velocities. The results of this procedure are illustrated in Fig. \ref{Figure9} for a compression protocol with parameters $v=5$cm/s and $\epsilon=0.6$, thus corresponding to the fastest process considered in Fig. \ref{Figure7}, in which disks and bars yield the same responsive force.

Fig. \ref{Figure9} considers a sequence of snapshots depicting different stages of the compression of samples consisting of disks (upper panel) and bars (lower panel). To represent the velocity of the particles in each snapshot, we employ a colored scale, in which the darkest color corresponds to zero velocity (i.e. stagnant zones), the blue-green color denotes particles moving practically at the piston velocity $v$, and with the red color representing the highest value obtained (which is approximately $2v$). Focusing on the case of the disks first, one notices that the external force applied to the system is primarily transmitted to particles closest to the piston, which move at nearly the compression speed $v$. These particles, in turn, convey part of the external force to their neighbors via the magnetic repulsion, which then pass it on to the next layer and so forth. As a result of this process, a network of force chains is created around the piston, highlighted in the upper panel of Fig. \ref{Figure9} by the colored area that ranges from greenish-blue to dark-blue. Note that particles that belong to this network move slower as they are farther away from the piston, displaying a dynamics that resembles a fingering flow pattern.

One observes that, just before the end of the compression process, the force-chain network extends over a region that covers roughly two-thirds of the remaining area, without affecting appreciably the particle distribution of the stagnant zones of the sample. After compression, in addition, two qualitatively different regions can be identified in the system, namely, a close packed domain with particles showing a seemingly regular spatial distribution (near the piston), and a diluted region of homogeneously distributed particles (away from the piston). These features are in good qualitative agreement with previous observations for the compressive response of repelling disks under variable initial conditions \cite{Modesto2022}, so that one expects that both the development of force chain networks, and the density profiles obtained at the end of a compression, are insensitive to the specific details of the fabrication protocol and system's history.

The above scenario is to be contrasted with that observed during the compression of the magnetic bars. In this case, the PIV reveals slightly larger velocities for the particles near the piston, thus suggesting smaller frictional contributions. More crucially, one also observes that the fingering (longer and thinner protrusions in the force chain network) penetrates deep into the samples during the compression and influences the particle distribution on the stagnant zones. As a result, we do not observe in this case the formation of two regions with markedly different densities, but a smooth density gradient, with the largest concentration of particles observed in the vicinity of the piston, and the minimum in the opposite limit of the sample. It is worth recalling here that, despite having such dissimilar particle distributions, the response force obtained at the end of compression is virtually the same for both samples.

\begin{figure}[ht!]
\begin{center}
 \includegraphics[width=\columnwidth]{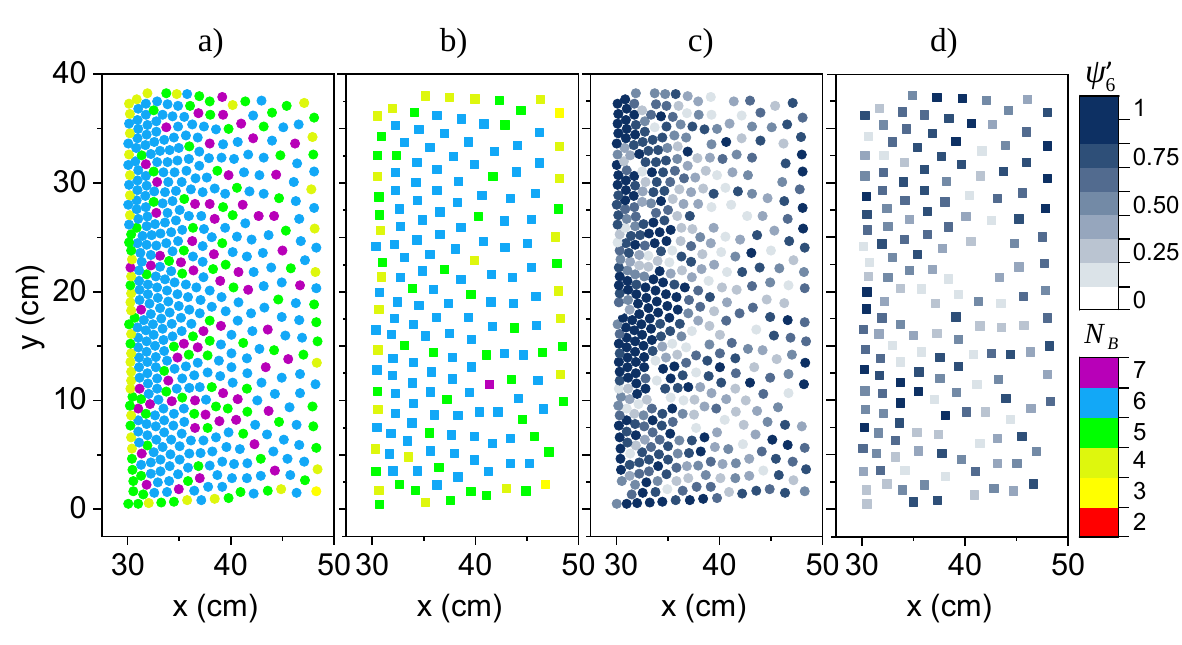}
 \caption{(a and b) Spatial diagrams indicating the number of nearest neighbors around the repelling disks (a, solid circles) and bars (b, solid squares) at the end of a compression protocol with parameters $v=5$ cm/s and $\epsilon=0.6$, represented by the color scale in the bottom right. (c and d) Corresponding values of bond orientational parameter $\psi_6^{\prime}$, represented by the palette of blue tones in the top right (compression along the x-axis in the horizontal XY plane).}
 \label{Figure10}
\end{center}
\end{figure}

To better understand the origin of this behavior, we also analyzed the spatial and orientational organization of the conforming particles at the end of each compression. The degree of spatial ordering in the system was determined via the so-called bond orientational parameter, defined as: 
\begin{equation}
\psi_6^{\prime} (\vec{r}) = \frac{1}{N_B} \sum_{ j=1}^{N_B} e^{i 6 \theta_j} 
\label{equation2}
\end{equation}
where $N_B$ is the number of nearest neighbors around a chosen arbitrary particle located at the position $\mathbf{r}$, and $0 \leq \theta_j<2 \pi$, with $\theta_j$ being the angle formed by the vector that localizes the $j$th-neighbor with respect to the selected particle, measured from an arbitrary reference axis. In terms of this order parameter, the hexagonal ordering in the present system is attained when $\psi_6^{\prime}\approx 1$. 

To compute $\psi_6^{\prime}$, we first determined the center of mass distribution of the particles, and a Delaunay triangulation was employed afterwards to estimate the number of neighbors for each particle in the samples. Fig. \ref{Figure10} illustrates the results of this procedure for disks (Fig. \ref{Figure10}a) and bars (Fig. \ref{Figure10}b), with the number of neighbors being represented by the colored scale displayed at the bottom right of this figure. In these images, the piston is located at $x=30$ cm from its initial position. Note that, in spite of the different particle numbers considered in each sample, and the distinct geometrical characteristics, in both cases one obtains mostly six nearest neighbors for each particle, but distributed over the whole area in a rather different way. 

Using this information and Eq. (\ref{equation2}), thus, we computed the bond orientational parameter $\psi_6^{\prime}$ in both systems. The results are shown in Figs. \ref{Figure10}c and \ref{Figure10}d for  samples composed by disks and bars, respectively, now using the palette of blue tones shown at the top right corner of the figure. These findings clearly highlight the hexagonal close packing distributions obtained in the vicinity of the piston for the case of the disks, represented by a much more darkened region on the left, which contrasts with the lighter area on the right that corresponds to a homogeneous distribution of particles. For the case of the bars, instead, no sign of hexagonal ordering is observed in all the space, but a uniform distribution of light blue squares that represents a homogeneous distribution of centers of mass.  Given the intrinsic geometry of the bar-shaped particles, it might be of interest to test other order parameters, such as $\psi_4^{\prime}$, more suitable to describe a squared lattice distribution of centers of mass. For the sake of brevity, however, we do not discuss this parameter here, but let us mention that its analysis reveals the same physical scenario as $\psi_6^{\prime}$, i.e., the absence of positional ordering. 

\begin{figure}[ht!]
\begin{center}
 \includegraphics[width=\columnwidth]{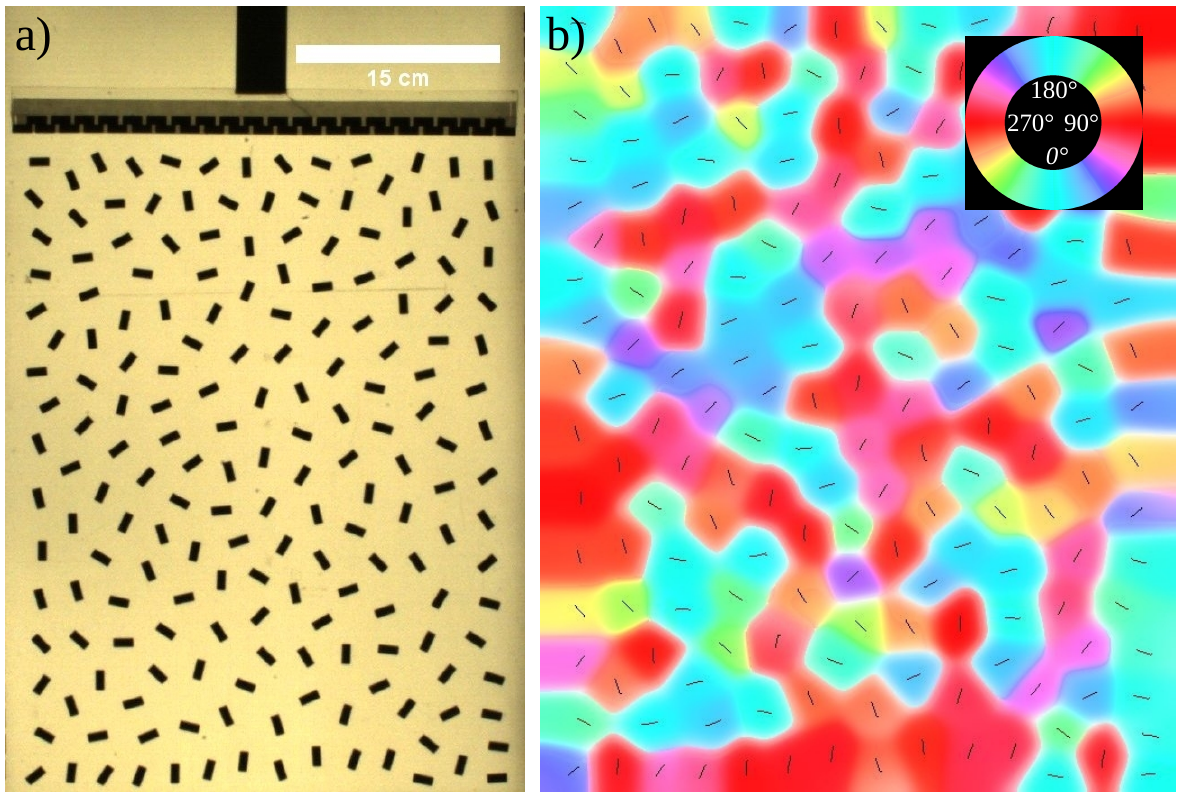}
 \caption{a) A initial random configuration of the granular bed composed of bars before compression, b) Colormap of the same image indicating the orientational distribution of the bars, with red corresponding to  alignment in the direction of the applied compression. The complete color scale depending on the angle of orientation respect to the piston is indicated in the inset.}
 \label{Figure11}
\end{center}
\end{figure}

The repelling bars, however, are prone to rotate during their compression, and hence, we also analyzed the distribution of orientations at the end of the compression. For this, we performed a qualitative analysis using the OrientationJ plugin from ImageJ, which allows to determine the orientation of the particles from a a binarized image of the system. Our procedure is briefly outlined in Fig. \ref{Figure11} for an arbitrary homogeneous and isotropic distribution of bar-shaped particles corresponding to an initial state. We start from an image of the system, such as that shown in Fig. \ref{Figure11}a, and process its binarized version in order to obtain a distribution of thinner bars (almost lines). We then employ the OrientationJ plugin, which assigns a color to the contour of an object as shown in Fig. \ref{Figure11}b with the corresponding color scale depicted on the inset. Hence, for the processed image of a rectangular bar, the red tones represent orientation in the direction of the compression,  blue tonalities represent orientation perpendicular to the direction of the piston stroke, and so on. 

\begin{figure}[ht!]
\begin{center}
 \includegraphics[width=\columnwidth]{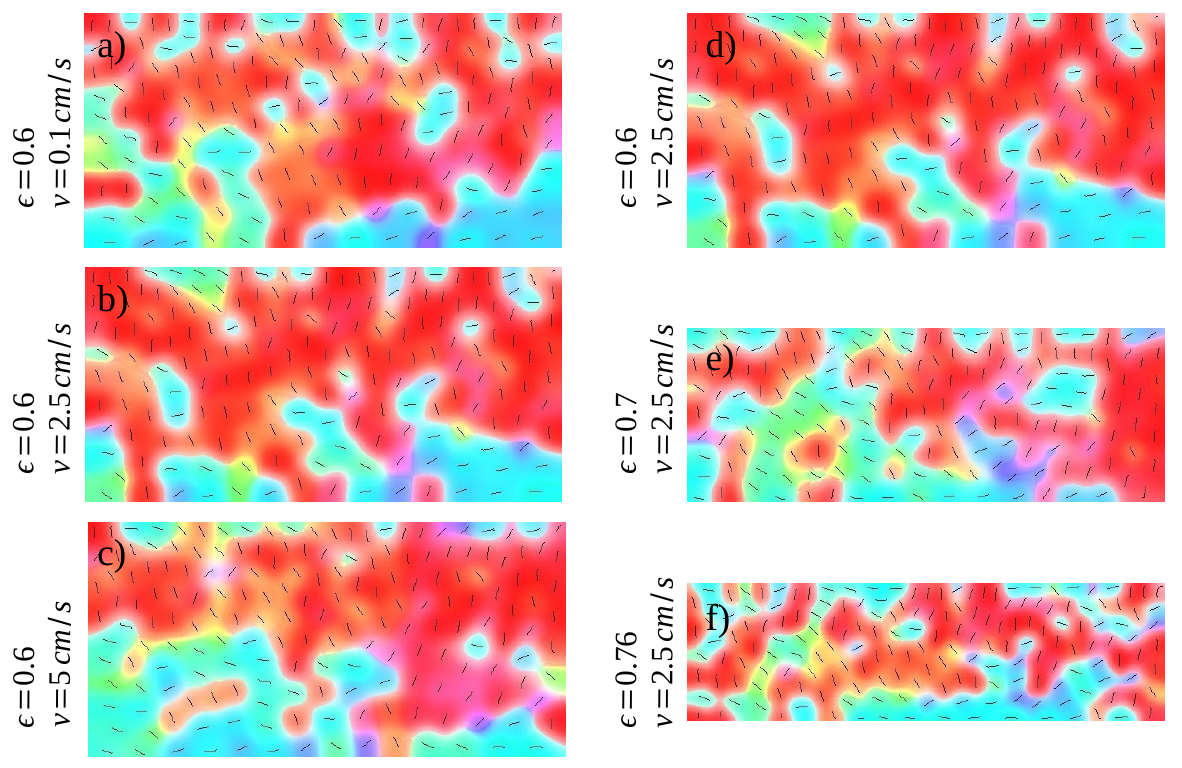}
 \caption{Orientational analysis of the final state of the granular bed after compression of the magnetic bars, for a-c) different compression speeds $v$ and d-f) different deformation ratios $\epsilon$. Colors are in correspondence with the color scale in Fig. \ref{Figure11}.}
 \label{Figure12}
\end{center}
\end{figure}

Using this method, we thus analyzed the orientational distributions obtained for the series of compressions previously considered. Our results are summarized in Fig. \ref{Figure12}. At fixed compaction ratio, $\epsilon=0.6$ (left column), the orientational ordering in the system at distinct compression rates $v$ is highlighted by the spread of red coloring within the sample, thus indicating that most of the bar-shaped particles tend to orient in the direction of compression. 

At first glance, we can notice that the largest region colored in red is obtained for the case $v=2.5$ cm/s. However, we observe that for faster compression, $v=5$ cm/s, one obtains almost two well separated color regions, namely, a larger red domain in the vicinity of the piston, and a smaller blue domain at the opposite boundary of the sample. Remarkably, this scenario is qualitatively comparable (to some extent) to that obtained for a sample of disks for the compression process ($\epsilon=0.6,v=5$ cm/s), illustrated above in Fig. \ref{Figure7}c and where the spatial ordering of the particles is observed in the vicinity of the piston, whereas the rest of the particles are homogeneously distributed in the rest of the system. 

Therefore, we observe that the response of both systems is mediated by the degree of particle ordering in the vicinity of the piston. Of course, in the case of particles with isotropic magnetic repulsion (disks), such ordering involves a hexagonal organization of magnets in the vicinity of the piston. For anisotropic repelling particles, instead, the underlying process mediating the response to compression is the orientational (nematic-like) ordering of the bars.

As a final relevant remark, let us comment that the samples of bars display a slightly different particle distribution for larger compaction ratios, which is markedly different behavior compared to the case of the disks. This is illustrated in the right column of Fig. \ref{Figure12}, for a series of compressions, at fixed $v=2.5$ cm/s but now for different compactation rations, as indicated. Clearly, the orientational ordering of the system decreases as the sample is more compressed, with no sign of an orientation region formed in the vicinity of the piston. This occurs because the magnets at the bottom wall (which are fixed horizontally) reorient the moving bars, altering the orientational distribution of the particles above this region. This suggests that the use of repelling particles with anisotropy in their magnetic repulsion allows to explore other additional mechanisms for the compressive response of the system in the high compression regime. Given the physical limitations of our Helle-Shaw cell, we leave the elucidation of these aspects for future work. 

\section{Concluding Remarks}\label{section5}

We have presented an experimental investigation of the collective behavior in a system of magnetic repelling particles under quasi two-dimensional confinement. By exploring different microscopic parameters and macroscopic conditions, such as particle shape, number concentration, degree of confinement and compression rate, we have outlined some of the most relevant characteristics of this system. Despite the non-contact nature of the interactions between particles, we found that the system exhibits static and dynamic behaviors that mimic those observed in conventional granular matter, driven by frictional contacts. 

First, we highlighted the ability of the repelling particles to stack, that is, to develop an angle of repose. This typical feature of granular matter arises from the balance between particles' weight and friction. As we discuss in our work, in our present system the frictional contributions are originated in the magnetic torques experienced by the constitutive particles, which are prone to flip due to the dipole-dipole interaction, but whose dipole inversion is constrained by the confining walls. 

The role of the geometrical shape of the conforming particles was emphasized by employing two kinds of magnets, namely, neodymium disks and bars, with the former displaying larger angles of repose with respect to the latter. At first glance, this feature might contrast with the case of ordinary granular matter, in which the opposite trend is normally obtained (i.e. elongated particles are expected to form steeper stags as compared to heaps formed with spherical grains). In the present case, however, the two-dimensional confinement promotes local hexagonal ordering for the repelling disks, which acts in favor of more stable piles. The rotational degrees of freedom of the bar-shaped particles, in contrast, hinder particle interlocking and promote flow, thus leading to smaller angles of repose.  

A second feature highlighted is the manifestation of the Janssen effect in this system of repelling particles, which results from the magnetic pressure induced by the gravitational-compaction that is redirected towards the confining walls. Again, we found a similar behavior as that observed in ordinary granular columns inside a container. By pouring gently the repelling particles, an increasingly larger fraction of the total weight of the added mass is redirected towards the rear and frontal panels of the confining cell. Above a certain height of a two-dimensional column formed with the repelling particles, thus, the magnetic torques generate enough friction against the glass plates of the cell, so that they can support the weight of the rest of the column. Remarkably, the average force profiles obtained in our measurements are well described by a Janssen-like expression, thus confirming the balance between vertical and transversal stresses in the magnetic medium. These findings reveal that the system of magnetic repelling particles can be described in some aspects as an effective continuum medium, as it is routinely done for conventional granular systems. 

To provide  insight into the microscopic mechanisms underlying the above behavior, we also analyzed the response under compression for both types of particles. Our measurements indicate that particle geometry plays a minor role in determining the compressive strength of the system. Instead, the number of particles and the total mass are the salient parameters, where an increase of the former reduces notably the amount of mass required to obtain a given resistance force, since it implies more contributions of non-contact repulsive forces in the system. Thus, in the interest of developing a magnetic granular damper, one should be focused in considering the largest possible amount of repelling particles, instead of focusing in the particle shape. 

The evolution of the spatial and orientational distributions of particles under compression also reveals interesting features. In the case of the disks, for example, an exponential growth of the compressive strength is produced by a gradual organization of the particles into a hexagonal structure, the extent of which increases with the the compression length. For the bars, on the other hand, no spatial ordering is observed upon compression. Instead, the same compressive response is the result of an orientational organization of the particles in the form of a locally nematic state, which grows in the vicinity of the compressing piston. We therefore conjecture that exploring additional mechanisms to enhance either hexagonal or nematic order, according to the case, could be a powerful tool to control the compressive response of a damping system based on confined magnets. Our results might provide useful insight in the development of such dampers, but also important, might be instrumental in the elucidation of more complex situations involving, for example, the use of different confining geometries, the simultaneous application of stresses in different directions, or the possibility to find the optimal size (or a mixture of sizes) to maximize the resistance to compression.
\backmatter

%
%
%

\bmhead{Acknowledgments}

The authors gratefully acknowledge financial support from CONACyT Mexico (now Secihti) and VIEP-BUAP project 2024.  M.A.G. acknowledges Ph.D. scholarship 1079091 ("Becas nacionales CONACyT 2023"). L.F.E.A. acknowledges the Consejo Nacional de Ciencia y Tecnolog{\'i}a (CONACYT, M{\'e}xico) for support through a Postdoctoral Fellowship (Grant No. I1200/224/2021). 

\section*{Declarations}

\begin{itemize}

\item Funding: Research supported by CONACYT Mexico through Frontier Science Project 140604 FORDECYT-PRONACES and VIEP-BUAP 2024-2025. \\

\item Conflict of interest: the authors have no relevant financial or non-financial interests to disclose.\\

\item Authors contributions: Experiments and data acquisition: M.A.G. Analysis and discussion: all authors. Supervision: F.P.V.  The initial draft was written by M.A.G. and L.F.E.A, and all authors revised, discussed and commented on previous versions of the manuscript.  All authors read and approved the final manuscript.

\end{itemize}

\bibliographystyle{sn-mathphys} 

\bibliography{references}

\providecommand{\noopsort}[1]{}\providecommand{\singleletter}[1]{#1}%

\begin{thebibliography}{47}
\ifx \bisbn   \undefined \def \bisbn  #1{ISBN #1}\fi
\ifx \binits  \undefined \def \binits#1{#1}\fi
\ifx \bauthor  \undefined \def \bauthor#1{#1}\fi
\ifx \batitle  \undefined \def \batitle#1{#1}\fi
\ifx \bjtitle  \undefined \def \bjtitle#1{#1}\fi
\ifx \bvolume  \undefined \def \bvolume#1{\textbf{#1}}\fi
\ifx \byear  \undefined \def \byear#1{#1}\fi
\ifx \bissue  \undefined \def \bissue#1{#1}\fi
\ifx \bfpage  \undefined \def \bfpage#1{#1}\fi
\ifx \blpage  \undefined \def \blpage #1{#1}\fi
\ifx \burl  \undefined \def \burl#1{\textsf{#1}}\fi
\ifx \doiurl  \undefined \def \doiurl#1{\url{https://doi.org/#1}}\fi
\ifx \betal  \undefined \def \betal{\textit{et al.}}\fi
\ifx \binstitute  \undefined \def \binstitute#1{#1}\fi
\ifx \binstitutionaled  \undefined \def \binstitutionaled#1{#1}\fi
\ifx \bctitle  \undefined \def \bctitle#1{#1}\fi
\ifx \beditor  \undefined \def \beditor#1{#1}\fi
\ifx \bpublisher  \undefined \def \bpublisher#1{#1}\fi
\ifx \bbtitle  \undefined \def \bbtitle#1{#1}\fi
\ifx \bedition  \undefined \def \bedition#1{#1}\fi
\ifx \bseriesno  \undefined \def \bseriesno#1{#1}\fi
\ifx \blocation  \undefined \def \blocation#1{#1}\fi
\ifx \bsertitle  \undefined \def \bsertitle#1{#1}\fi
\ifx \bsnm \undefined \def \bsnm#1{#1}\fi
\ifx \bsuffix \undefined \def \bsuffix#1{#1}\fi
\ifx \bparticle \undefined \def \bparticle#1{#1}\fi
\ifx \barticle \undefined \def \barticle#1{#1}\fi
\bibcommenthead
\ifx \bconfdate \undefined \def \bconfdate #1{#1}\fi
\ifx \botherref \undefined \def \botherref #1{#1}\fi
\ifx \url \undefined \def \url#1{\textsf{#1}}\fi
\ifx \bchapter \undefined \def \bchapter#1{#1}\fi
\ifx \bbook \undefined \def \bbook#1{#1}\fi
\ifx \bcomment \undefined \def \bcomment#1{#1}\fi
\ifx \oauthor \undefined \def \oauthor#1{#1}\fi
\ifx \citeauthoryear \undefined \def \citeauthoryear#1{#1}\fi
\ifx \endbibitem  \undefined \def \endbibitem {}\fi
\ifx \bconflocation  \undefined \def \bconflocation#1{#1}\fi
\ifx \arxivurl  \undefined \def \arxivurl#1{\textsf{#1}}\fi
\csname PreBibitemsHook\endcsname

\bibitem{varela}
\begin{barticle}
\bauthor{\bsnm{Varela-Rosales}, \binits{N.R.}},
\bauthor{\bsnm{Santarossa}, \binits{A.}},
\bauthor{\bsnm{Engel}, \binits{M.}},
\bauthor{\bsnm{P{\"o}schel}, \binits{T.}}:
\batitle{Granular binary mixtures improve energy dissipation efficiency of
  granular dampers}.
\bjtitle{Granular Matter}
\bvolume{25}(\bissue{3}),
\bfpage{49}
(\byear{2023})
\end{barticle}
\endbibitem

\bibitem{braj}
\begin{barticle}
\bauthor{\bsnm{Prasad}, \binits{B.B.}},
\bauthor{\bsnm{Duvigneau}, \binits{F.}},
\bauthor{\bsnm{Juhre}, \binits{D.}},
\bauthor{\bsnm{Woschke}, \binits{E.}}:
\batitle{Damping performance of particle dampers with different granular
  materials and their mixtures}.
\bjtitle{Applied Acoustics}
\bvolume{200},
\bfpage{109059}
(\byear{2022})
\end{barticle}
\endbibitem

\bibitem{terzioglu1}
\begin{barticle}
\bauthor{\bsnm{Terzioglu}, \binits{F.}},
\bauthor{\bsnm{Rongong}, \binits{J.A.}},
\bauthor{\bsnm{Lord}, \binits{C.E.}}:
\batitle{Motional phase maps for estimating the effectiveness of granular
  dampers}.
\bjtitle{Mechanical Systems and Signal Processing}
\bvolume{188},
\bfpage{110038}
(\byear{2023})
\end{barticle}
\endbibitem

\bibitem{hamzeh}
\begin{barticle}
\bauthor{\bsnm{Pourtavakoli}, \binits{H.}},
\bauthor{\bsnm{Parteli}, \binits{E.J.}},
\bauthor{\bsnm{P{\"o}schel}, \binits{T.}}:
\batitle{Granular dampers: does particle shape matter?}
\bjtitle{New Journal of Physics}
\bvolume{18}(\bissue{7}),
\bfpage{073049}
(\byear{2016})
\end{barticle}
\endbibitem

\bibitem{terzioglu2}
\begin{barticle}
\bauthor{\bsnm{Terzioglu}, \binits{F.}},
\bauthor{\bsnm{Rongong}, \binits{J.A.}},
\bauthor{\bsnm{Lord}, \binits{C.E.}}:
\batitle{Influence of particle sphericity on granular dampers operating in the
  bouncing bed motional phase}.
\bjtitle{Journal of Sound and Vibration}
\bvolume{554},
\bfpage{117690}
(\byear{2023})
\end{barticle}
\endbibitem

\bibitem{lu}
\begin{barticle}
\bauthor{\bsnm{Lu}, \binits{Z.}},
\bauthor{\bsnm{Chen}, \binits{X.}},
\bauthor{\bsnm{Zhang}, \binits{D.}},
\bauthor{\bsnm{Dai}, \binits{K.}}:
\batitle{Experimental and analytical study on the performance of particle tuned
  mass dampers under seismic excitation}.
\bjtitle{Earthquake Engineering \& Structural Dynamics}
\bvolume{46}(\bissue{5}),
\bfpage{697}--\blpage{714}
(\byear{2017})
\end{barticle}
\endbibitem

\bibitem{Pacheco2013}
\begin{barticle}
\bauthor{\bsnm{Pacheco-V{\'a}zquez}, \binits{F.}},
\bauthor{\bsnm{Dorbolo}, \binits{S.}}:
\batitle{Rebound of a confined granular material: combination of a bouncing
  ball and a granular damper}.
\bjtitle{Scientific reports}
\bvolume{3}(\bissue{1}),
\bfpage{2158}
(\byear{2013})
\end{barticle}
\endbibitem

\bibitem{Sanchez2012}
\begin{barticle}
\bauthor{\bsnm{S{\'a}nchez}, \binits{M.}},
\bauthor{\bsnm{Rosenthal}, \binits{G.}},
\bauthor{\bsnm{Pugnaloni}, \binits{L.A.}}:
\batitle{Universal response of optimal granular damping devices}.
\bjtitle{Journal of Sound and Vibration}
\bvolume{331}(\bissue{20}),
\bfpage{4389}--\blpage{4394}
(\byear{2012})
\end{barticle}
\endbibitem

\bibitem{rozenblat}
\begin{barticle}
\bauthor{\bsnm{Rozenblat}, \binits{Y.}},
\bauthor{\bsnm{Portnikov}, \binits{D.}},
\bauthor{\bsnm{Levy}, \binits{A.}},
\bauthor{\bsnm{Kalman}, \binits{H.}},
\bauthor{\bsnm{Aman}, \binits{S.}},
\bauthor{\bsnm{Tomas}, \binits{J.}}:
\batitle{Strength distribution of particles under compression}.
\bjtitle{Powder Technology}
\bvolume{208}(\bissue{1}),
\bfpage{215}--\blpage{224}
(\byear{2011})
\end{barticle}
\endbibitem

\bibitem{Azadeh2001}
\begin{barticle}
\bauthor{\bsnm{Samadani}, \binits{A.}},
\bauthor{\bsnm{Kudrolli}, \binits{A.}}:
\batitle{Angle of repose and segregation in cohesive granular matter}.
\bjtitle{Physical Review E}
\bvolume{64}(\bissue{5}),
\bfpage{051301}
(\byear{2001})
\end{barticle}
\endbibitem

\bibitem{Jaeger1989}
\begin{barticle}
\bauthor{\bsnm{Jaeger}, \binits{H.}},
\bauthor{\bsnm{Liu}, \binits{C.-h.}},
\bauthor{\bsnm{Nagel}, \binits{S.R.}}:
\batitle{Relaxation at the angle of repose}.
\bjtitle{Physical review letters}
\bvolume{62}(\bissue{1}),
\bfpage{40}
(\byear{1989})
\end{barticle}
\endbibitem

\bibitem{degennes}
\begin{barticle}
\bauthor{\bsnm{Boutreux}, \binits{T.}},
\bauthor{\bsnm{Rapha{\"e}l}, \binits{E.}},
\bauthor{\bsnm{De~Gennes}, \binits{P.}}:
\batitle{Propagation of a pressure step in a granular material: The role of
  wall friction}.
\bjtitle{Physical Review E}
\bvolume{55}(\bissue{5}),
\bfpage{5759}
(\byear{1997})
\end{barticle}
\endbibitem

\bibitem{Pacheco2021}
\begin{barticle}
\bauthor{\bsnm{Pacheco-V{\'a}zquez}, \binits{F.}},
\bauthor{\bsnm{Omura}, \binits{T.}},
\bauthor{\bsnm{Katsuragi}, \binits{H.}}:
\batitle{Undulating compression and multistage relaxation in a granular column
  consisting of dust particles or glass beads}.
\bjtitle{Physical Review Research}
\bvolume{3}(\bissue{1}),
\bfpage{013190}
(\byear{2021})
\end{barticle}
\endbibitem

\bibitem{Beverloo1961}
\begin{barticle}
\bauthor{\bsnm{Beverloo}, \binits{W.A.}},
\bauthor{\bsnm{Leniger}, \binits{H.A.}},
\bauthor{\bparticle{Van~de} \bsnm{Velde}, \binits{J.}}:
\batitle{The flow of granular solids through orifices}.
\bjtitle{Chemical engineering science}
\bvolume{15}(\bissue{3-4}),
\bfpage{260}--\blpage{269}
(\byear{1961})
\end{barticle}
\endbibitem

\bibitem{Hilton2011}
\begin{barticle}
\bauthor{\bsnm{Hilton}, \binits{J.}},
\bauthor{\bsnm{Cleary}, \binits{P.}}:
\batitle{Granular flow during hopper discharge}.
\bjtitle{Physical Review E—Statistical, Nonlinear, and Soft Matter Physics}
\bvolume{84}(\bissue{1}),
\bfpage{011307}
(\byear{2011})
\end{barticle}
\endbibitem

\bibitem{Alvaro2012}
\begin{barticle}
\bauthor{\bsnm{Janda}, \binits{A.}},
\bauthor{\bsnm{Zuriguel}, \binits{I.}},
\bauthor{\bsnm{Maza}, \binits{D.}}:
\batitle{Flow rate of particles through apertures obtained from self-similar
  density<? format?> and velocity profiles}.
\bjtitle{Physical review letters}
\bvolume{108}(\bissue{24}),
\bfpage{248001}
(\byear{2012})
\end{barticle}
\endbibitem

\bibitem{Rubio2015}
\begin{barticle}
\bauthor{\bsnm{Rubio-Largo}, \binits{S.M.}},
\bauthor{\bsnm{Janda}, \binits{A.}},
\bauthor{\bsnm{Maza}, \binits{D.}},
\bauthor{\bsnm{Zuriguel}, \binits{I.}},
\bauthor{\bsnm{Hidalgo}, \binits{R.}}:
\batitle{Disentangling the free-fall arch paradox in silo discharge}.
\bjtitle{Physical review letters}
\bvolume{114}(\bissue{23}),
\bfpage{238002}
(\byear{2015})
\end{barticle}
\endbibitem

\bibitem{Arean2020}
\begin{barticle}
\bauthor{\bsnm{Are{\'a}n}, \binits{M.}},
\bauthor{\bsnm{Boschan}, \binits{A.}},
\bauthor{\bsnm{Cachile}, \binits{M.A.}},
\bauthor{\bsnm{Aguirre}, \binits{M.A.}}:
\batitle{Granular flow through an aperture: Influence of obstacles near the
  outlet}.
\bjtitle{Physical Review E}
\bvolume{101}(\bissue{2}),
\bfpage{022901}
(\byear{2020})
\end{barticle}
\endbibitem

\bibitem{opsomer}
\begin{barticle}
\bauthor{\bsnm{Opsomer}, \binits{E.}},
\bauthor{\bsnm{Merminod}, \binits{S.}},
\bauthor{\bsnm{Schockmel}, \binits{J.}},
\bauthor{\bsnm{Vandewalle}, \binits{N.}},
\bauthor{\bsnm{Berhanu}, \binits{M.}},
\bauthor{\bsnm{Falcon}, \binits{E.}}:
\batitle{Patterns in magnetic granular media at the crossover from two to three
  dimensions}.
\bjtitle{Physical Review E}
\bvolume{102}(\bissue{4}),
\bfpage{042907}
(\byear{2020})
\end{barticle}
\endbibitem

\bibitem{thorens}
\begin{barticle}
\bauthor{\bsnm{Thorens}, \binits{L.}},
\bauthor{\bsnm{M{\aa}l{\o}y}, \binits{K.J.}},
\bauthor{\bsnm{Bourgoin}, \binits{M.}},
\bauthor{\bsnm{Santucci}, \binits{S.}}:
\batitle{Magnetic janssen effect}.
\bjtitle{Nature Communications}
\bvolume{12}(\bissue{1}),
\bfpage{2486}
(\byear{2021})
\end{barticle}
\endbibitem

\bibitem{Pacheco2015}
\begin{barticle}
\bauthor{\bsnm{Lumay}, \binits{G.}},
\bauthor{\bsnm{Schockmel}, \binits{J.}},
\bauthor{\bsnm{Hen{\'a}ndez-Enr{\'\i}quez}, \binits{D.}},
\bauthor{\bsnm{Dorbolo}, \binits{S.}},
\bauthor{\bsnm{Vandewalle}, \binits{N.}},
\bauthor{\bsnm{Pacheco-Vazquez}, \binits{F.}}:
\batitle{Flow of magnetic repelling grains in a two-dimensional silo}.
\bjtitle{Papers in physics}
\bvolume{7}(\bissue{2}),
\bfpage{0}--\blpage{0}
(\byear{2015})
\end{barticle}
\endbibitem

\bibitem{lumay}
\begin{barticle}
\bauthor{\bsnm{Lumay}, \binits{G.}},
\bauthor{\bsnm{Vandewalle}, \binits{N.}}:
\batitle{Controlled flow of smart powders}.
\bjtitle{Physical Review E—Statistical, Nonlinear, and Soft Matter Physics}
\bvolume{78}(\bissue{6}),
\bfpage{061302}
(\byear{2008})
\end{barticle}
\endbibitem

\bibitem{Modesto2022}
\begin{barticle}
\bauthor{\bsnm{Modesto}, \binits{J.}},
\bauthor{\bsnm{Dorbolo}, \binits{S.}},
\bauthor{\bsnm{Katsuragi}, \binits{H.}},
\bauthor{\bsnm{Pacheco-V{\'a}zquez}, \binits{F.}},
\bauthor{\bsnm{Sobral}, \binits{Y.D.}}:
\batitle{Experimental and numerical investigation of the compression and
  expansion of a granular bed of repelling magnetic disks}.
\bjtitle{Granular Matter}
\bvolume{24}(\bissue{4}),
\bfpage{105}
(\byear{2022})
\end{barticle}
\endbibitem

\bibitem{Cox}
\begin{barticle}
\bauthor{\bsnm{Cox}, \binits{M.}},
\bauthor{\bsnm{Wang}, \binits{D.}},
\bauthor{\bsnm{Bar{\'e}s}, \binits{J.}},
\bauthor{\bsnm{Behringer}, \binits{R.P.}}:
\batitle{Self-organized magnetic particles to tune the mechanical behavior of a
  granular system}.
\bjtitle{Europhysics letters}
\bvolume{115}(\bissue{6}),
\bfpage{64003}
(\byear{2016})
\end{barticle}
\endbibitem

\bibitem{tsuchikusa}
\begin{botherref}
\oauthor{\bsnm{Tsuchikusa}, \binits{K.}},
\oauthor{\bsnm{Yamamoto}, \binits{K.}},
\oauthor{\bsnm{Katsura}, \binits{M.}},
\oauthor{\bparticle{de} \bsnm{Paula}, \binits{C.}},
\oauthor{\bsnm{Modesto}, \binits{J.}},
\oauthor{\bsnm{Dorbolo}, \binits{S.}},
\oauthor{\bsnm{Pacheco-V{\'a}zquez}, \binits{F.}},
\oauthor{\bsnm{Sobral}, \binits{Y.}},
\oauthor{\bsnm{Katsuragi}, \binits{H.}}:
Disordering two-dimensional magnet-particle configurations using bidispersity.
The Journal of Chemical Physics
\textbf{158}(21)
(2023)
\end{botherref}
\endbibitem

\bibitem{Lopez2023}
\begin{barticle}
\bauthor{\bsnm{L{\'o}pez-Gonz{\'a}lez}, \binits{F.}},
\bauthor{\bsnm{Pacheco-V{\'a}zquez}, \binits{F.}},
\bauthor{\bsnm{Donado}, \binits{F.}}:
\batitle{Ordering of a granular layer of cubes under strain-induced shear and
  vibration}.
\bjtitle{Physica A: Statistical Mechanics and its Applications}
\bvolume{620},
\bfpage{128768}
(\byear{2023})
\end{barticle}
\endbibitem

\bibitem{disks}
\begin{botherref}
Neodymium disks.
\url{https://www.imanes.com.mx/discos_de_neodimio/118-iman-de-neodimio-disco-0197-x-0118.html}.
Accessed: April 15, 2024
\end{botherref}
\endbibitem

\bibitem{bars}
\begin{botherref}
Neodymium bars.
\url{https://www.imanes.com.mx/blocks-de-neodimio/65-iman-de-neodimio-block-0600-x-0260-x-0110.html}.
Accessed: April 15, 2024
\end{botherref}
\endbibitem

\bibitem{hamza2018}
\begin{barticle}
\bauthor{\bsnm{Al-Hashemi}, \binits{H.M.B.}},
\bauthor{\bsnm{Al-Amoudi}, \binits{O.S.B.}}:
\batitle{A review on the angle of repose of granular materials}.
\bjtitle{Powder technology}
\bvolume{330},
\bfpage{397}--\blpage{417}
(\byear{2018})
\end{barticle}
\endbibitem

\bibitem{elekes}
\begin{barticle}
\bauthor{\bsnm{Elekes}, \binits{F.}},
\bauthor{\bsnm{Parteli}, \binits{E.J.}}:
\batitle{An expression for the angle of repose of dry cohesive granular
  materials on earth and in planetary environments}.
\bjtitle{Proceedings of the National Academy of Sciences}
\bvolume{118}(\bissue{38}),
\bfpage{2107965118}
(\byear{2021})
\end{barticle}
\endbibitem

\bibitem{khanal}
\begin{barticle}
\bauthor{\bsnm{Khanal}, \binits{M.}},
\bauthor{\bsnm{Elmouttie}, \binits{M.}},
\bauthor{\bsnm{Adhikary}, \binits{D.}}:
\batitle{Effects of particle shapes to achieve angle of repose and force
  displacement behaviour on granular assembly}.
\bjtitle{Advanced Powder Technology}
\bvolume{28}(\bissue{8}),
\bfpage{1972}--\blpage{1976}
(\byear{2017})
\end{barticle}
\endbibitem

\bibitem{sperl2}
\begin{barticle}
\bauthor{\bsnm{Sperl}, \binits{M.}}:
\batitle{Experiments on corn pressure in silo cells--translation and comment of
  janssen's paper from 1895}.
\bjtitle{Granular Matter}
\bvolume{8}(\bissue{2}),
\bfpage{59}--\blpage{65}
(\byear{2006})
\end{barticle}
\endbibitem

\bibitem{Janssen}
\begin{barticle}
\bauthor{\bsnm{Janssen}, \binits{H.}}:
\batitle{Investigations of pressure of grain in silo}.
\bjtitle{Vereins Eutscher Ingenieure Zeitschrift}
\bvolume{39},
\bfpage{1045}--\blpage{1049}
(\byear{1895})
\end{barticle}
\endbibitem

\bibitem{hagen}
\begin{barticle}
\bauthor{\bsnm{Hagen}, \binits{G.H.L.}}
\bjtitle{Annalen der Physik und Chemie}
\bvolume{46},
\bfpage{423}
(\byear{1839})
\end{barticle}
\endbibitem

\bibitem{sperl1}
\begin{barticle}
\bauthor{\bsnm{Tighe}, \binits{B.P.}},
\bauthor{\bsnm{Sperl}, \binits{M.}}:
\batitle{Pressure and motion of dry sand: translation of hagen's paper from
  1852}.
\bjtitle{Granular Matter}
\bvolume{9},
\bfpage{141}--\blpage{144}
(\byear{2007})
\end{barticle}
\endbibitem

\bibitem{Vanel1999}
\begin{barticle}
\bauthor{\bsnm{Vanel}, \binits{L.}},
\bauthor{\bsnm{Cl{\'e}ment}, \binits{E.}}:
\batitle{Pressure screening and fluctuations at the bottom of a granular
  column}.
\bjtitle{The European Physical Journal B-Condensed Matter and Complex Systems}
\bvolume{11},
\bfpage{525}--\blpage{533}
(\byear{1999})
\end{barticle}
\endbibitem

\bibitem{Vanel2000}
\begin{barticle}
\bauthor{\bsnm{Vanel}, \binits{L.}},
\bauthor{\bsnm{Claudin}, \binits{P.}},
\bauthor{\bsnm{Bouchaud}, \binits{J.-P.}},
\bauthor{\bsnm{Cates}, \binits{M.}},
\bauthor{\bsnm{Cl{\'e}ment}, \binits{E.}},
\bauthor{\bsnm{Wittmer}, \binits{J.}}:
\batitle{Stresses in silos: comparison between theoretical models and new
  experiments}.
\bjtitle{Physical review letters}
\bvolume{84}(\bissue{7}),
\bfpage{1439}
(\byear{2000})
\end{barticle}
\endbibitem

\bibitem{Bratberg2005}
\begin{barticle}
\bauthor{\bsnm{Bratberg}, \binits{I.}},
\bauthor{\bsnm{M{\aa}l{\o}y}, \binits{K.}},
\bauthor{\bsnm{Hansen}, \binits{A.}}:
\batitle{Validity of the janssen law in narrow granular columns}.
\bjtitle{The European Physical Journal E}
\bvolume{18},
\bfpage{245}--\blpage{252}
(\byear{2005})
\end{barticle}
\endbibitem

\bibitem{Huang2016}
\begin{barticle}
\bauthor{\bsnm{Huang}, \binits{Y.}},
\bauthor{\bsnm{Daniels}, \binits{K.E.}}:
\batitle{Friction and pressure-dependence of force chain communities in
  granular materials}.
\bjtitle{Granular Matter}
\bvolume{18}(\bissue{4}),
\bfpage{85}
(\byear{2016})
\end{barticle}
\endbibitem

\bibitem{Katsuragi2016}
\begin{bbook}
\bauthor{\bsnm{Katsuragi}, \binits{H.}}, \betal:
\bbtitle{Physics of Soft Impact and Cratering}.
\bpublisher{Springer}, \blocation{???}
(\byear{2016})
\end{bbook}
\endbibitem

\bibitem{Yann2003}
\begin{barticle}
\bauthor{\bsnm{Bertho}, \binits{Y.}},
\bauthor{\bsnm{Giorgiutti-Dauphin{\'e}}, \binits{F.}},
\bauthor{\bsnm{Hulin}, \binits{J.-P.}}:
\batitle{Dynamical janssen effect on granular packing with moving walls}.
\bjtitle{Physical review letters}
\bvolume{90}(\bissue{14}),
\bfpage{144301}
(\byear{2003})
\end{barticle}
\endbibitem

\bibitem{Windows2019}
\begin{barticle}
\bauthor{\bsnm{Windows-Yule}, \binits{C.}},
\bauthor{\bsnm{M{\"u}hlbauer}, \binits{S.}},
\bauthor{\bsnm{Cisneros}, \binits{L.T.}},
\bauthor{\bsnm{Nair}, \binits{P.}},
\bauthor{\bsnm{Marzulli}, \binits{V.}},
\bauthor{\bsnm{P{\"o}schel}, \binits{T.}}:
\batitle{Janssen effect in dynamic particulate systems}.
\bjtitle{Physical Review E}
\bvolume{100}(\bissue{2}),
\bfpage{022902}
(\byear{2019})
\end{barticle}
\endbibitem

\bibitem{Karim2014}
\begin{barticle}
\bauthor{\bsnm{Karim}, \binits{M.Y.}},
\bauthor{\bsnm{Corwin}, \binits{E.I.}}:
\batitle{Eliminating friction with friction: 2d janssen effect in a
  friction-driven system}.
\bjtitle{Physical Review Letters}
\bvolume{112}(\bissue{18}),
\bfpage{188001}
(\byear{2014})
\end{barticle}
\endbibitem

\bibitem{deGennes1999}
\begin{barticle}
\bauthor{\bparticle{de} \bsnm{Gennes}, \binits{P.-G.}}:
\batitle{Granular matter: a tentative view}.
\bjtitle{Reviews of modern physics}
\bvolume{71}(\bissue{2}),
\bfpage{374}
(\byear{1999})
\end{barticle}
\endbibitem

\bibitem{Pacheco2021-2}
\begin{bchapter}
\bauthor{\bsnm{Pacheco-V{\'a}zquez}, \binits{F.}},
\bauthor{\bsnm{Omura}, \binits{T.}},
\bauthor{\bsnm{Katsuragi}, \binits{H.}}:
\bctitle{Grain size effect on the compression and relaxation of a granular
  column: solid particles vs dust agglomerates}.
In: \bbtitle{EPJ Web of Conferences},
vol. \bseriesno{249},
p. \bfpage{07005}
(\byear{2021}).
\bcomment{EDP Sciences}
\end{bchapter}
\endbibitem

\bibitem{Valdes2012}
\begin{barticle}
\bauthor{\bsnm{Valdes}, \binits{J.R.}},
\bauthor{\bsnm{Fernandes}, \binits{F.L.}},
\bauthor{\bsnm{Einav}, \binits{I.}}:
\batitle{Periodic propagation of localized compaction in a brittle granular
  material}.
\bjtitle{Granular Matter}
\bvolume{14},
\bfpage{71}--\blpage{76}
(\byear{2012})
\end{barticle}
\endbibitem

\bibitem{Guillard2015}
\begin{barticle}
\bauthor{\bsnm{Guillard}, \binits{F.}},
\bauthor{\bsnm{Golshan}, \binits{P.}},
\bauthor{\bsnm{Shen}, \binits{L.}},
\bauthor{\bsnm{Valdes}, \binits{J.R.}},
\bauthor{\bsnm{Einav}, \binits{I.}}:
\batitle{Dynamic patterns of compaction in brittle porous media}.
\bjtitle{Nature Physics}
\bvolume{11}(\bissue{10}),
\bfpage{835}--\blpage{838}
(\byear{2015})
\end{barticle}
\endbibitem

\end{thebibliography}

\end{document}